\documentclass[aps,twocolumn,superscriptaddress,floatfix]{revtex4-1}
\usepackage{graphicx,amsfonts,amssymb,amsmath,mathrsfs,hyperref,bm,dsfont}

\newif\ifhyper
\hypertrue
\ifhyper
\hypersetup{
   citecolor = {red},
   colorlinks = {true}, % false
   linkcolor = {blue},
   urlcolor = {blue} % magenta
}
\fi

\def\be{\begin{equation}}
\def\ee{\end{equation}}
\def\bea{\begin{eqnarray}}
\def\eea{\end{eqnarray}}

\newcommand{\Zd}{\mathbb{Z}_2}

\newcommand{\expectval}[1]{\langle #1\rangle}

\begin{document}

%\title{Tensor Network simulation of the 3D Kitaev materials}
%\title{3D Kitaev spin liquids via tensor network: ground state and thermodynamic properties}
\title{Thermodynamics of 3D Kitaev quantum spin liquids via tensor networks}

\author{Saeed S. Jahromi}
\email{saeed.jahromi@dipc.org}
\affiliation{Donostia International Physics Center, Paseo Manuel de Lardizabal 4, E-20018 San Sebasti\'an, Spain}

\author{Hadi Yarloo}
\affiliation{Department of Physics, Sharif University of Technology, P.O.Box 11155-9161, Tehran, Iran}

\author{Rom\'an Or\'us}
\affiliation{Donostia International Physics Center, Paseo Manuel de Lardizabal 4, E-20018 San Sebasti\'an, Spain}
\affiliation{Ikerbasque Foundation for Science, Maria Diaz de Haro 3, E-48013 Bilbao, Spain}
\affiliation{Multiverse Computing, Paseo de Miram\'on 170, 20014 San Sebasti\'an, Spain}

\begin{abstract}
We study the 3D Kitaev and Kitaev-Heisenberg models respectively on the hyperhoneycomb and hyperoctagon lattices, both at zero and finite-temperature, in the thermodynamic limit. Our analysis relies on advanced tensor network (TN) simulations based on graph Projected Entangled-Pair States (gPEPS). We map out the TN phase diagrams of the models and characterize their underlying gapped and gapless phases both at zero and finite temperature. In particular, we demonstrate how cooling down the hyperhoneycomb system from high-temperature leads to  fractionalization of spins to itinerant Majorana fermions and gauge fields that occurs in two separate temperature regimes, leaving their fingerprint on specific heat as a double-peak feature as well as on other quantities such as the thermal entropy, spin-spin correlations and bond entropy. Using the Majorana representation of the Kitaev model, we further show that the low-temperature thermal transition to the Kitaev quantum spin liquid (QSL) phase is associated with the non-trivial Majorana band topology and the presence of Weyl nodes, which manifests itself via non-vanishing Chern number and finite thermal Hall conductivity. Beyond the pure Kitaev limit, we study the 3D Kitaev-Heisenberg (KH) model on the hyperoctagon lattice and extract the full phase diagram for different Heisenberg couplings. We further explore the thermodynamic properties of the magnetically-ordered regions in the KH model and show that, in contrast to the QSL phase, here the thermal phase transition follows the standard Landau symmetry-breaking theory.  
       
\end{abstract}

\maketitle

\section{Introduction}
 Quantum spin liquids (QSL) \cite{Savary2017} are distinct phases of matter with exotic properties such as long-range entanglement, topological order \cite{Levin2006,Kitaev2006,Jahromi2017} and fractionalized excitations \cite{Kitaev2003,Levin2005}. Different instances of QSLs have already been observed in different settings such as quantum antiferromagnets \cite{Balents2010,Liao2017,Poilblanc2019,Jahromi2018a}, superconducting phases \cite{Anderson1987,Poilblanc2014} and topologically ordered spin systems \cite{Wen1990,Wen1995,Levin2005,Kitaev2003,Kitaev2006,Jahromi2013a,Jahromi2013,Jahromi2017,Jahromi2016,Mohseninia2015,Capponi2014}.

The two-dimensional (2D) Kitaev model on the honeycomb lattice \cite{Kitaev2006} is one of the famous examples of a QSL which has played a major role in a deeper understanding of the physics of quantum phases of matter, both theoretically and experimentally. The Kitaev model is a quantum spin system with anisotropic Ising-like bond directional interactions that naturally arises as an interplay of crystal-field and strong spin-orbit coupling in a variety of $4d$ and $5d$ materials \cite{Khaliullin2005,Trebst2017,Pesin2010}. Due to the bond-directional interactions, the Kitaev model on trivalent lattices is highly frustrated with dominant quantum fluctuations. At zero temperature, the ground state of these Mott insulators forms a highly entangled QSL in which the original spin degrees of freedom are fractionalized into itinerant (non-interacting) Majorana fermions and an emergent static $\Zd$ gauge field. The resulting QSL is a gapless state which becomes gapped by breaking time-reversal symmetry (TRS) \cite{Kitaev2006}. This so-called Kitaev spin liquid is a topologically ordered state that is known to host non-abelian Ising anyons, gapped flux excitations (visions) \cite{Kitaev2006,Kitaev2003}, and a chiral gapless Majorana edge mode which gives rise to a quantized thermal quantum Hall effect (QHE) at low-temperature regime \cite{Kasahara2018}.

\begin{figure}  
\centerline{\includegraphics[width=6cm]{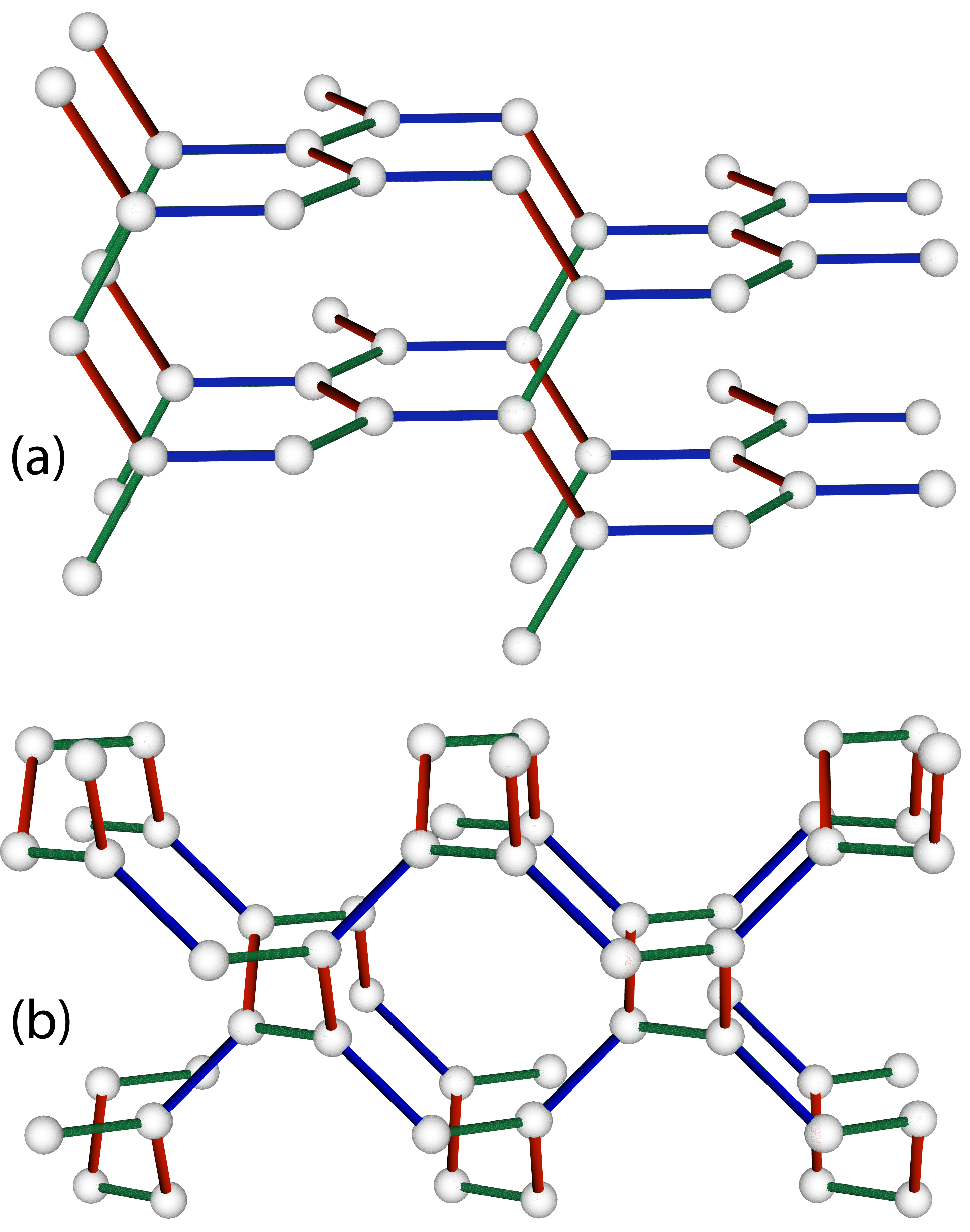}}
\caption{(Color online) Kitaev model on the (a) hyperhoneycomb and (b) hyperoctagon lattices. The red, green and blue links denote the Ising interactions of type $x, y$ and $z$.}
\label{Fig:lattice}
\end{figure}

The exact solvability of the Kitaev model remains valid for 3D trivalent lattices \cite{OBrien2016,Mandal2009,SI2008428}. This has largely motivated the study of the Kitaev model on lattices such as the hyperhoneycomb \cite{Hermanns2015,Takayama2015} and hyperoctagon \cite{Hermanns2014} (see Fig.~\ref{Fig:lattice} and also Ref.~\cite{OBrien2016} for a full list of relevant 3D Kitaev lattices) which are extensions of the honeycomb \cite{Kitaev2006} and square-octagon \cite{Kargarian2010} lattices to 3D. Most importantly, the recent discovery of a 3D material with strong bond directional spin-orbit interactions in $\beta-\rm{Li_2IrO_3}$ and $\gamma-\rm{Li_2IrO_3}$ compounds \cite{Takayama2015,Modic2014} has attracted considerable attention to the theoretical and experimental study of 3D Kitaev materials. Similar to the 2D Kitaev honeycomb model, the low-energy physics of 3D Kitaev QSLs is also described as a gapless Majorana metal in the background of $\Zd$ gauge fields \cite{OBrien2016}. However, depending on the underlying lattice geometries, the Fermi surfaces of the Majorana metals are described as topological Majorana Fermi surfaces, nodal lines, or Weyl points \cite{OBrien2016,Hermanns2014,Hermanns2015}. 

Away from the exactly solvable point or at finite temperature, the analytical tractability of the Kitaev model becomes highly non-trivial or impossible. In such situations, ground-state properties of the model can only be studied by advanced numerical techniques. For example, studying the phase diagram of the Kitaev model in the presence of the Heisenberg interaction, which naturally arises as the next-leading interaction in the spin-orbit Mott insulators \cite{Chaloupka2010,Chaloupka2013}, is highly challenging with analytical techniques. Besides, studying the thermodynamic properties of the Kitaev QSL on different 3D lattices at finite temperatures is numerically challenging. While the 2D Kitaev model on different lattices has been studied largely by state-of-the-art numerical methods such as exact diagonalization (ED) \cite{Morita2020,Hickey2019,Koga2018,SUZUKI2018637}, quantum Monte Carlo (QMC) \cite{Nasu2015}, and tensor network (TN) algorithms \cite{Czarnik2019}, the study of a generic Kitaev model on different 3D lattices is only limited to mean-field treatment \cite{Lee2014}, series expansion \cite{Singh2017}, and more recently a QMC which remains sign-free as long as the gauge fields are static and the Majorana representation remains valid \cite{Eschmann2020,Mishchenko2017,Nasu2014,Nasu2014a,Nasu2015}.

Although TN methods have been shown to be one of the most promising techniques for accurate simulation of the 2D strongly correlated systems both at zero- \cite{Corboz2014a,Corboz2012a,Corboz2013,Corboz2014,Jahromi2018,Jahromi2018a,Schmoll2020,Sadrzadeh2016,Jahromi2020,Jahromi2019} and finite-temperature \cite{Wietek2019,Kshetrimayum2019,Qu2019,Czarnik2012,Czarnik2015a,Czarnik2015,Czarnik2019a,Kshetrimayum2017,Verstraete2004a,Jahromi2020a}, their application to 3D lattices, and in particular at finite temperature, has largely been left behind mostly due to technical challenges. It is therefore crucial to develop new efficient tools to simulate generic 3D quantum many-body systems that are not directly tractable by, say, QMC methods. In this paper, we use our recent graph-based infinite projected entangled-pair state algorithm (gPEPS) \cite{Jahromi2019} to study the ground-state properties of the 3D Kitaev model in the thermodynamic limit. More specifically, we simulate the spin-$1/2$ Kitaev model, i.e., Eq.\eqref{eq:H-Kitaev} on the hyperhoneycomb lattice, computing their zero temperature phase diagram. Next, we use a variant of our TN algorithm for calculating the thermal density matrix of infinite-size quantum systems, the so-called thermal gPEPS (TgPEPS) \cite{Jahromi2020} and study the thermodynamic properties of the systems at finite temperature. We show that the TgPEPS can faithfully capture the intermediate-to-high temperature regimes in the thermodynamic limit, which are the most relevant ones in experimental probes of Kitaev materials \cite{Eschmann2020}. 

We particularly demonstrate how fractionalization of the original spin degrees of freedom to Majorana fermions and gauge fields leaves its fingerprint on the local observables such as nearest-neighbor correlations and bond entanglement. In order to crosscheck and supplement our TN simulations, we use the Majorana representation of the Kitaev model \cite{Kitaev2006} and extract the thermodynamic properties of the system for the hyperhoneycomb lattice particularly at very low-temperatures, which is the challenging regime for TN algorithms. We further calculate the Chern number and thermal Hall conductivity of the Kitaev model and capture the thermal phase transition beyond which the gauge degrees of freedom are stabilized in the background and the ground state ends up being a highly entangled QSL. Away from the exactly-solvable point, we study the 3D Kitaev-Heisenberg (KH) model, i.e., Hamiltonian \eqref{eq:H-KH}, in the hyperoctagon lattice, and extract the full phase diagram of the KH model in different regimes of the Heisenberg couplings. We show how different phases and phase boundaries can be identified by measuring different quantities such as magnetization, entanglement entropy, and spin-spin correlations. Our study is further complemented by exploring the thermodynamic properties of different magnetically ordered regions in the phase diagram of the KH model.     

The paper is organized as follows: In Sec.~\ref{sec:model} we introduce the Kitaev Hamiltonian and review the ground state properties of the model on different 3D lattice structures. In Sec.~\ref{sec:method} we discuss the details of our TN algorithm for simulating both ground state and the thermal density matrix of local Hamiltonians on any arbitrary lattice structure. Next in Sec.~\ref{sec:zero_T_pure} we discuss the TN phase diagram of the pure 3D Kitaev hyperhoneycomb model at zero-temperature. The thermodynamic properties of the Kitaev QSL at finite-temperature are discussed in Sec.~\ref{sec:fin_T_pure}, and the 3D Kitaev-Heisenberg hyperoctagon model is studied in Sec.~\ref{sec:kitaev_heisenberg}. Finally, Sec.~\ref{sec:conclude} is devoted to the discussion and conclusions.

\section{Model}  \label{sec:model}  

The Kitaev model was first introduced by Alexei Kitaev on the 2D honeycomb lattice in the context of topological quantum computation \cite{Kitaev2006}. The intriguing properties of the model soon attracted huge attention from the condensed matter perspective since it was suggested that the ground state of the system is a QSL \cite{Savary2017}. The Hamiltonian of the Kitaev model is given by  
\be
\mathcal{H}_{\rm Kitaev}=\sum_{\langle i,j \rangle,\gamma} K_{\gamma}\, S_i^{\gamma} S_j^{\gamma},
\label{eq:H-Kitaev}
\ee
where the sum runs over three subclasses of bonds labeled by $\gamma=x,y,z$, denoting the three Ising-like interactions on the corresponding link. While the original Kitaev model was introduced using spin-$1/2$ Pauli operators, our definition here is based on generic spin operators. 

The Kitaev model is exactly solvable on trivalent lattices, i.e., lattices with three links connected to each site. Using the local transformation $ {S_i^\gamma=\frac{i}{2} b^\gamma_i c_i} $, the original spin operators can be represented by four Majorana fermions. The original interacting spin model takes then the bilinear form, 
\bea
\label{eq:H-Majorana}
\mathcal{H}_{\rm mf}&=&\dfrac{i}{4} \sum_{ \langle i,j \rangle} A_{ij}  c_i c_j,\\
A_{ij}&=& \dfrac{K_{\gamma}}{2} u^\gamma_{ij}, \nonumber 
\eea
where ${\hat{u}^\gamma_{ij}=-\hat{u}^\gamma_{ji}=ib_i^\gamma b_j^\gamma}$ are bond operators obtained from grouping Majoranas ${b^\gamma}$ along nearest-neighbor links ${\langle i,j \rangle}$. The bond operators commute with each other and with the Hamiltonian \eqref{eq:H-Majorana}. They are, therefore, conserved integrals of motion which square to identity and their eigenvalues are given by $u_{ij}=\pm1$. Within this parton construction, the new degrees of freedom are now the non-interacting Majorana fermions $c_i$ coupled to a static $\Zd$ gauge field $u_{ij}$.

Given a fixed gauge configuration $\{u_{ij}\}$, the spectrum of the system can be readily extracted by diagonalizing the skew-symmetry matrix $iA$. Its eigenvalues come in pairs $ \pm\epsilon_\mu $ as a result of the inherent particle-hole symmetry of the (single-particle) Majorana Hamiltonian \eqref{eq:H-Majorana}. The ground state of the system is thus given by a configuration of gauge fields which minimizes the energy. According to Lieb's theorem, depending on the lattice structure, its symmetry, and dimensionality, the ground state is generically given by $0$- or $\pi$-flux configuration of gauge fields. 

While Lieb's theorem is valid for 2D trivalent lattices, its extension to 3D trivalent structures is hampered since the symmetry requirement for the theorem is not fulfilled in general, except for hyperhoneycomb lattice. One may therefore use numerically exact QMC simulations to unambiguously determine the ground-state gauge pattern. A careful analysis in this regard shows that all bipartite trivalent 3D lattices still follow the general conclusions of Lieb's theorem (in spite of not being applicable) in the absence of additional geometrical constraints leading to "gauge frustration" (see Refs.~\cite{Eschmann2020} for details).

%Since the $\Zd$ flux excitations of the Kitaev model, or visons, are always gapped, the low-energy physics of the model can be controlled by its underlying Majorana band structure. 
The zero-temperature phase diagram of the Kitaev model on the 2D honeycomb lattice is known to host two QSL phases. The first one is gapless and emerges near the isotropic point of Hamiltonian~\eqref{eq:H-Kitaev}, i.e., ${K\equiv K_x=K_y=K_z}$. The second one is actually three equivalent gapped phases that arise whenever one of the couplings dominates the other two. The low-energy effective theory of the three gapped phases is given by an abelian topological phase known as the toric code \cite{Kitaev2003}. The gapless phase is also known to become gapped by applying time-reversal symmetry (TRS) breaking perturbations to the Hamiltonian~\eqref{eq:H-Kitaev} such as a uniform magnetic field, $-\sum_i \vec{B}.\, S_i$ along the $[111]$ direction. When the field strength $B$ is small compared to the vison gap $ \Delta $, one can derive a low-energy effective model using third-order perturbation theory in the ground-state flux sector, yielding
\be
H_{eff}=\sum_{\langle i,j \rangle,\gamma} K_\gamma S_i^{\gamma} S_j^{\gamma} - {\kappa} \sum_{i,j,k} \, S^\alpha_i S^\beta_j S^\gamma_k,
\label{eq:H-Kitaev-Field}
\ee
where the three-spin coupling constant $\kappa\sim B_xB_yB_z/\Delta^2$ encodes the strength of the effective magnetic field, and triples $i,j,k$ indicate three neighbouring sites with strictly different bond-type $\alpha$, $\beta$ and $ \gamma $.

In the effective Majorana representation~\eqref{eq:H-Kitaev-Field}, the three-spin interaction turns into a next-nearest-neighbour hopping term between sites $i$ and $k$ connected by site $j$. In the 2D Kitaev models, such a hopping term  opens a gap at the Dirac points of the bulk spectrum and yields a chiral edge mode with a non-vanishing Chern number $\nu$ (other third-order terms are irrelevant to the Dirac-gap opening in the renormalization group sense). The resulting gapped state is $\Zd$ topologically ordered with non-abelian anyon excitations.
 
In the low-temperature regime, the thermal Hall conductance of the 2D Kitaev model shows a quantized value with respect to the field strength $ \kappa $, i.e., $\mathcal{I}_{xy}= \frac{\pi}{12} \nu T$, which is exactly half of the two-dimensional thermal Hall conductance in the integer QHE. Such a half-integer quantization is a signature of a topologically-protected chiral edge mode of charge-neutral Majorana fermions, which have half the degrees of freedom of conventional fermions. 

The Kitaev model on 3D lattices still shares some of the properties of the 2D honeycomb version, including  a rich variety of gapless $\Zd$ QSL phases. The main difference with the 2D version though is that, according to the symmetry classification of free-fermion systems~\cite{Altland1997}, breaking the TRS will not give rise to topologically protected gapped phases for 3D Kitaev models. Despite not being topological in a strict sense, these models exhibit however a finite but non-quantized thermal quantum Hall effect in the presence of a magnetic field (see Sec.~\ref{sec:fin_T_pure}). Moreover, the nodal structure of the Majorana fermions on different 3D Kitaev lattices have distinct properties such as nodal lines with a flat surface band, topologically protected Weyl points with the so-called surface Fermi arcs, and a Majorana semi-metal with a Majorana Fermi surface characterized by a finite Majorana density of states at lowest-lying energy levels \cite{OBrien2016}. 

In this paper we focus on two specific examples of 3D Kitaev structures, namely the hyperhoneycomb and hyperoctagon lattices, shown in Fig.~\ref{Fig:lattice}. Both lattices are among the most studied examples of 3D Kitaev materials. While the hyperoctagon lattice is the 3D extension of the square-octagon lattice, the hyperhoneycomb is the natural extension of the honeycomb lattice to 3D. The latter has a experimental realization in ${\beta-\mathrm{Li_2IrO_3}}$ compounds \cite{Takayama2015,Modic2014}. Besides, both hyperhoneycomb and hyperoctagon lattices contain two independent types of loop operators of length $10$ around their plaquettes which guarantee a ground state with vanishing total flux \cite{OBrien2016,Eschmann2020}. 

While the phase diagram of both lattices is composed of a gapless region around the isotropic point surrounded by three gapped phases at the corners (see Fig.~\ref{Fig:3Dphasediag}), they have different Majorana Fermi surfaces. The gapless region of the hyperoctagon phase diagram harbours two distinct Majorana Fermi surface regions: a trivial one and another region with topological protected Fermi surface enclosing a Weyl node at finite energy. In contrast, the gapless modes in the bulk spectrum of the hyperhoneycomb lattice form a closed line of Dirac nodes which is protected by TRS. Breaking the TRS will gap out the Majorana Fermi line, leaving pairs of Weyl points in the bulk, exactly at zero energy, and gapless Fermi arcs on the surface \cite{Hermanns2015}. 

\section{Methods} 
\label{sec:method} 

Let us now briefly sketch how TNs can be used to obtain the phase diagram and thermodynamic properties of the Kitaev model on the aforementioned 3D lattices at zero and finite-temperature.

\subsection{Tensor network algorithms for 3D lattices}

TN methods \cite{Orus2014,Orus2014a,Orus2019,Ran2017,Biamonte2017,Verstraete2008} have played a major role in the discovery of many exotic phases of matter in recent years. The basic idea at work in these methods is to write the wave function of local quantum many-body (QMB) Hamiltonians in a very efficient way based on their entanglement structure \cite{Orus2014}. Matrix product states (MPS), which provide a natural TN representation of 1D QMB systems, are probably the most famous example of TN states, since they are at the core of well-known algorithms such as density matrix renormalization group (DMRG) \cite{White1993,White2004}. Another well-known example is that of projected entangled-pair state (PEPS) \cite{Verstraete2006,Verstraete2008,Orus2014,Orus2014a} and its variant for infinite lattices, i.e, infinite-PEPS (iPEPS).  

The TN representation for the ground state of a local lattice Hamiltonian can generally be obtained by variational methods \cite{Corboz2016} or approaches based on evolution in imaginary-time \cite{Orus2009,Phien2015,Vidal2007}. In spite of their great promise and success, these techniques are restricted due to the implementation challenges. Examples of this are the lattice geometry as well as the efficient approximation of environments \cite{Levin2007,Orus2009}.  During past years tremendous progress has been put forward to simulate different 2D lattice structures with PEPS. However, there have been not so many examples of 3D simulations with TNs. Thanks to our recent  graph-based iPEPS algorithm (which we call gPEPS) \cite{Jahromi2019}, we have managed to simulate local quantum lattice models in arbitrary dimensions and lattice geometries. 

The gPEPS algorithm relies on the so-called {\it structure-matrix} (SM) construction, which resolves the geometrical implementation challenges \cite{Jahromi2019, Jahromi2020a}. The SM provides a compact way for storing the connectivity information of the underlying TN, i.e., each column of the SM corresponds to one of the links of the lattice and contains all the details about the neighbouring tensors, their interconnecting indices, and their bond dimensions. One can then fully automatize the TN update by looping over the columns of the SM in a very systematic way, without the burden of complications due to geometry (see Refs.~\cite{Jahromi2019} for detailed discussions and examples of SM for different lattices). The ground state of local nearest-neighbour Hamiltonians is further approximated by using the simple-update (SU) technique based on imaginary-time evolution (ITE). Within the SU scheme, the environments of local tensors are given by bond matrices obtained from local singular-value decompositions (SVD), which provide a mean-field-like approximation to the environment and correlations around local tensors. While more accurate techniques such as full-update (FU) \cite{Corboz2012a,Corboz2013,Corboz2014} have also been designed to capture the full correlation in the system, their implementation and truncation is largely limited by lattice geometry particularly for 3D lattices. Nevertheless, the mean-field approximation of the environment in the SU has been shown to be very good for higher-dimensional systems as well as thermal states, in turn making the SU a quite accurate option in these situations. 

\begin{figure}  
\centerline{\includegraphics[width=7cm]{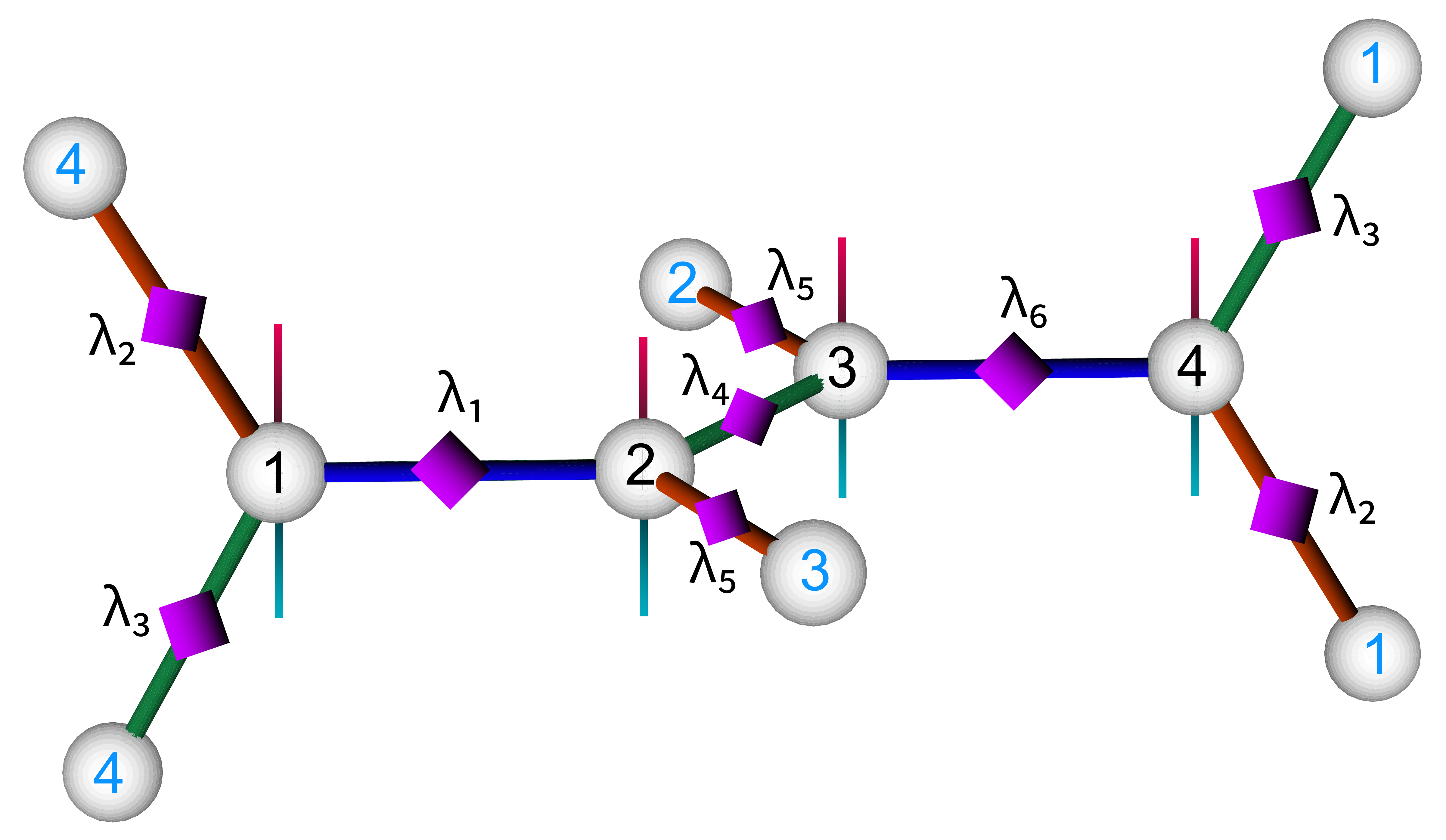}}
\caption{(Color online) Thermal density matrix PEPO of a four-site unit-cell of the hyperhoneycomb lattice. The vertical red and blue legs of tensors denote the bra and ket physical degrees of freedom for each local site.}
\label{Fig:10_3_b_TN}
\end{figure}

In order to study the thermodynamic properties of QMB systems at finite-temperature we target the thermal density-matrix (TDM) of the corresponding Hamiltonian $H$, i.e., ${\rho_\beta=e^{-\beta H}}$, ${\beta=1/T}$ being the inverse temperature. The TDM of the system (see Fig.~\ref{Fig:10_3_b_TN} for the TDM of the hyperhoneycomb lattice) can then be approximated by evolving in imaginary-time for a time ${\beta/2}$ both the bra and ket degrees of freedom starting from the maximally-entangled infinite temperature state, i.e., ${\rho_\beta = e^{-\beta H/2} \cdot {\mathbb I} \cdot e^{-\beta H/2}}$. Similar to the ground state simulation, our thermal gPEPS algorithm (TgPEPS) \cite{Jahromi2020a} uses both SU and mean-field environment for local optimization of the TDM in arbitrary dimensions. The expectation values of local operators and correlators can then be calculated as ${\langle \hat{\mathcal{O}} \rangle_\beta ={\rm Tr}(\rho_\beta \hat{\mathcal{O}})/{\rm Tr}(\rho_\beta)}$ where, ${\rm Tr}$ denotes the trace operation (see Ref.~\cite{Jahromi2020a} for more details on the method).

In order to simulate the 3D Kitaev model on the hyperhoneycomb and hyperoctagon lattices, we used translationally invariant unit-cells of $32$-sites. Our SU optimization technique for both gPEPS and TgPEPS was based on imaginary-time evolution accompanied by proper truncations at the level of local SVDs. Our algorithms have further been stabilized by proper choice of gauge-fixing and super-orthogonalization of local tensors \cite{Jahromi2019}. 

%\subsection{Berry curvature and thermal Hall effect}
%
%%\emph{\textbf{Berry curvature and thermal Hall effect}---.}
%While the phase diagram of the 3D Kitaev model on both
%hyperhoneycomb and hyperoctagon lattices are pretty similar,
%they have distinct Fermi surfaces with different topological
%properties at the isotropic point of the gapless region. As we
%already pointed out, our TN simulations yield the ground state
%wave function in real space. Therefore our TN techniques are
%blind to momentum dependant characterization of the 3D Kitaev QSLs.  However, for completeness of the study and as a complement to our tensor network simulations, we used the Majorana representation of the Kitaev model and calculated the Chern number and Hall conductivity of the 3D
%Kitaev model on hyperhoneycomb an hyperoctagon lattices both at zero and finite-temperature.
%These quantities will further serve as a powerful probe for finite temperature characterization of the Kitaev QSLs, which will be discussed in future sections in more details. We refer the interested reader to Ref. [] for technical details on the analytical calculation of the Chern number using the Majorana representations of the Kitaev model. Details about the calculation on the thermal Hall conductivity can further be find in Ref. [].

\section{Hyperhoneycomb Kitaev Spin-Liquids}

Let us start by presenting our results for the 3D Kitaev model in the hyperhoneycomb lattice at zero temperature. Results for the hyperoctagon lattice are very similar nd can be obtained analogously. 

\subsection{$T=0$ phase diagram} 
\label{sec:zero_T_pure}  

%\subsection{Tensor Network Phase Diagram}
\begin{figure}  
\centerline{\includegraphics[width=\columnwidth]{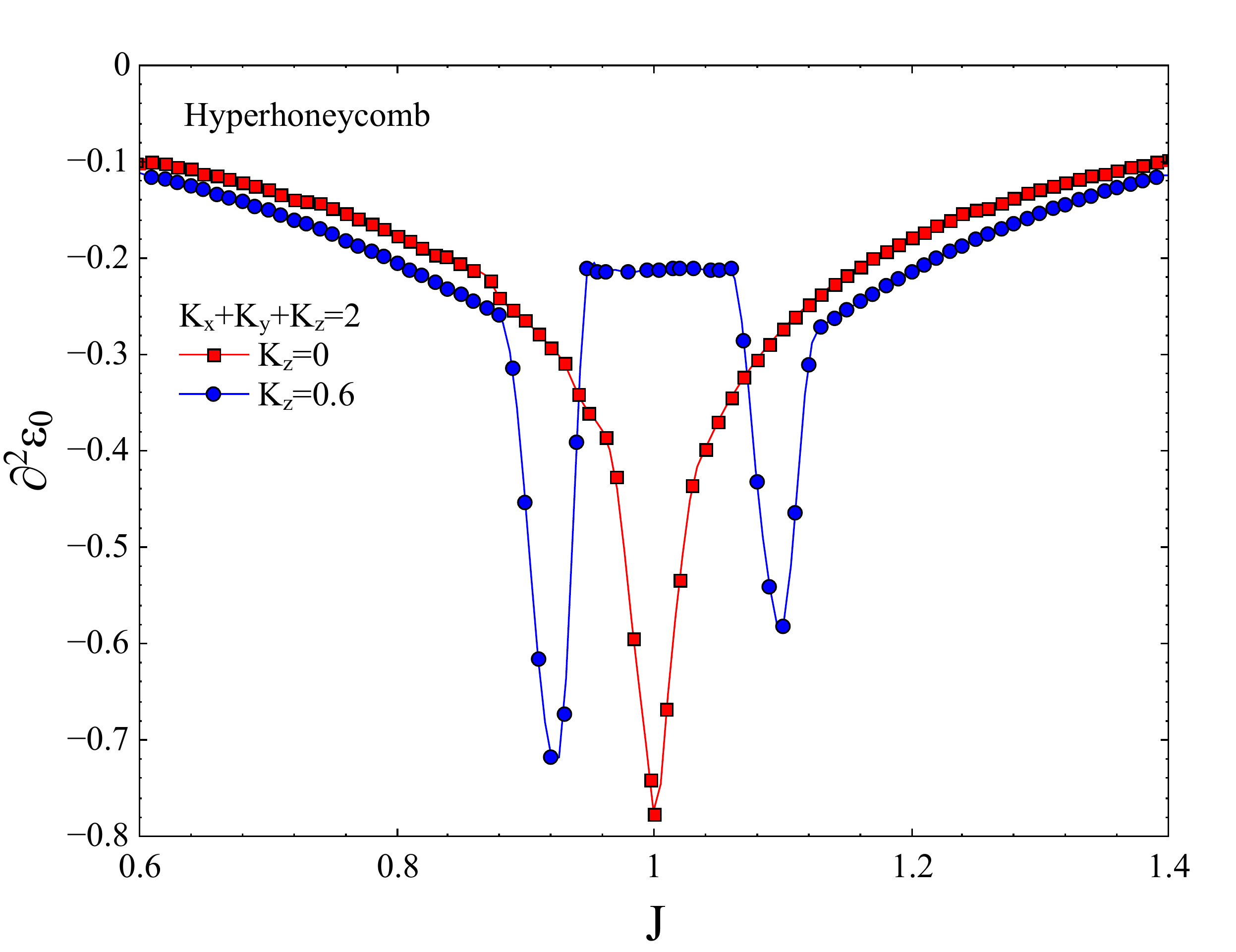}}
\caption{(Color online) Second-derivative of the ground state energy of the Kitaev model on the hyperhoneycomb lattice along the scan-lines shown in Fig.~\ref{Fig:3Dphasediag}. The vertical axis $J$ shows the points  along the line ${K_x+K_y=2-K_z}$ for ${K_z=0, 0.6}$. A similar picture holds for the hyperoctagon lattice. The small asymmetry in the curve for ${K_z=0, 0.6}$ is due to small numerical inaccuracies in the 2nd-derivative.}
\label{Fig:D2EHH}
\end{figure}
\begin{figure}  
\centerline{\includegraphics[width=\columnwidth]{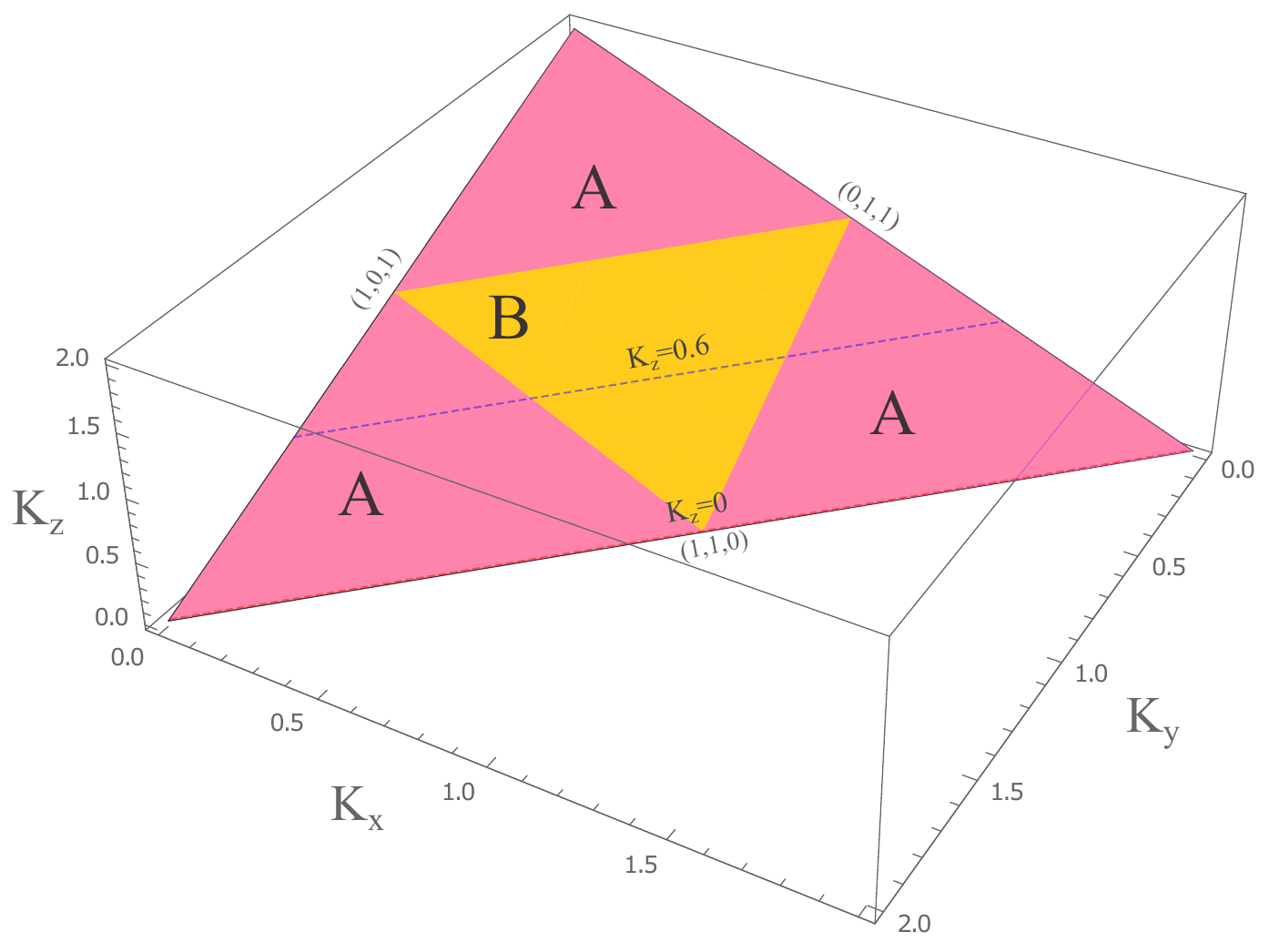}}
\caption{(Color online) TN phase diagram of the Kitaev model on the hyperhoneycomb lattice at zero temperature on the $K_x+K_y+K_z=2$ plane. The phase diagram is composed of one gapless spin liquid in center surrounded by three gapped phases at corners which arise when one of the coupling dominates the two others. The phase boundaries are second-order quantum phase transitions captured by second derivative of the ground state energy (see Fig.~\ref{Fig:D2EHH}) along scan lines. Two examples of these scan lines have been shown at $K_z=0, 0.6$. A similar picture holds for the hyperoctagon lattice.}
\label{Fig:3Dphasediag}
\end{figure}

The full phase diagram of the 3D Kitaev model on both Hyperhoneycomb and Hyperoctagon lattices pretty much looks the same as Fig.~\ref{Fig:3Dphasediag} which is composed of a gapless phase surrounded by three gapped regions. The phase boundaries can be captured analytically by locating the location of the closure of the gap of the spectrum of the quadratic Majorana Hamiltonian \eqref{eq:H-Majorana} in the momentum space \cite{Kitaev2006}. In the TN representation, we only have access to the ground state wave function of the 3D Kitaev model in the real space. The phase boundaries, however, can still be located by different physical observables and their derivatives such as the 2nd-derivative of the ground state energy per-site $\varepsilon_0$, as shown in Fig.~\ref{Fig:D2EHH} for the hyperhoneycomb lattice. The two curves show the behaviour of $\partial^2_J\varepsilon_0$ along scan lines $J$ with fixed $K_z=0,0.6$ in the $K_x+K_y+K_z=2$ plane (dashed lines in Fig.~\ref{Fig:3Dphasediag}). The 2nd-derivatives of the energy show sharp discontinuities when crossing a phase boundary, being this a clear signal of second-order quantum phase transitions. 

\subsection{Kitaev QSL at finite-$T$}
\label{sec:fin_T_pure}

\begin{figure}  
	\centerline{\includegraphics[width=\columnwidth]{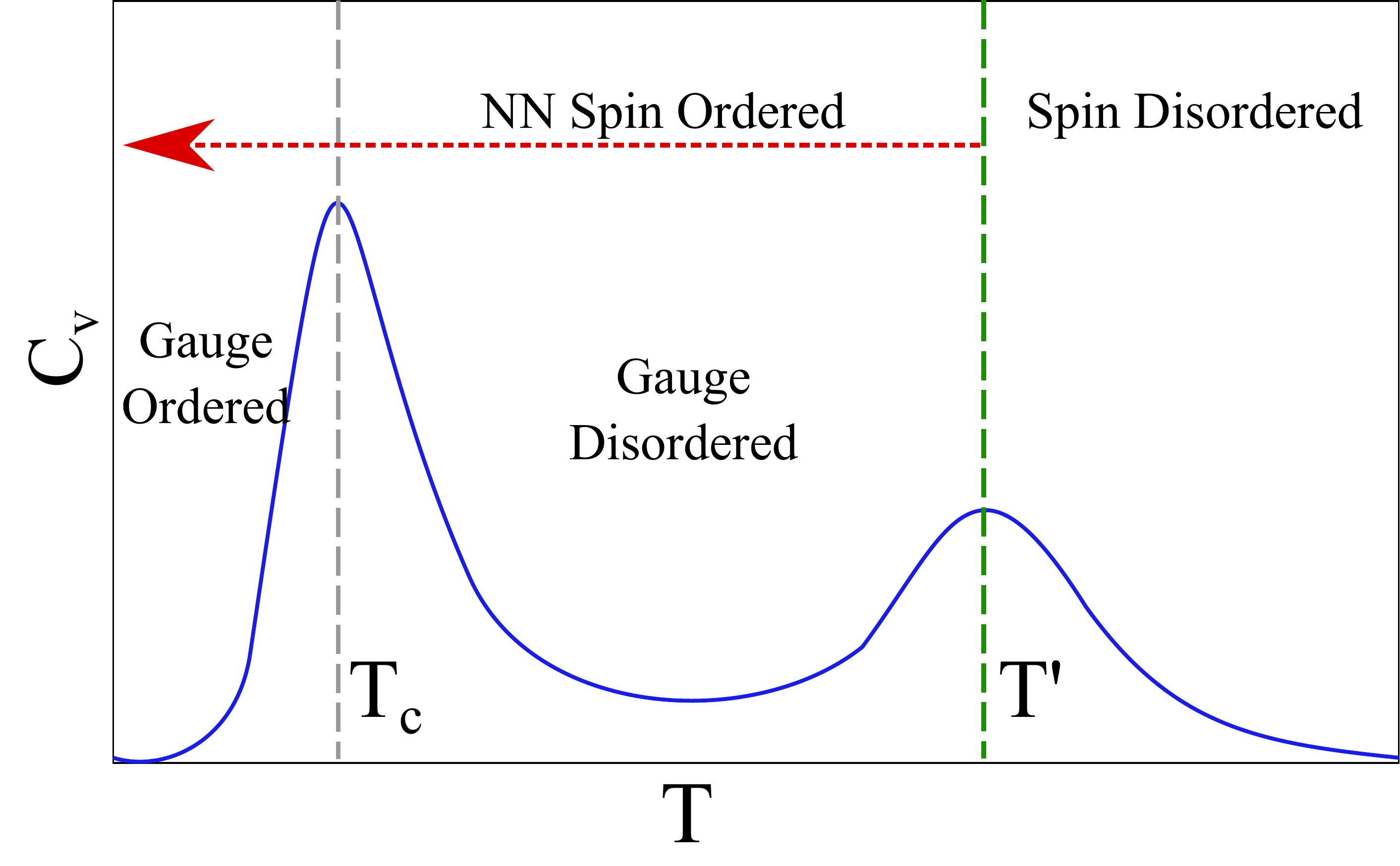}}
	\caption{(Color online) Schematic diagram of the specific heat $C_v$ of the generic Kitaev model showing a double-peak feature versus temperature. The high-temperature peak occurs at a crossover temperature $T'$ below which the spins are fractionalized to Majorana fermions and gauge field. The low-temperature peak reveals the thermal phase transition below which the gauge fields become ordered.}
	\label{Fig:Double_peak_sketch}
\end{figure}

It has already been known that the specific heat $C_v$ of the Kitaev model on generic lattices exhibits a double-peak behaviour \cite{Nasu2014,Nasu2014a,Nasu2015} at two different temperature regimes. One peak corresponds to the order of the Majorana bandwidth at $T'\sim K$, and the other  is correlated with the size of the vison gap, typically of the order $T_c\sim K/100$ (see the schematic diagram of Fig.~\ref{Fig:Double_peak_sketch}). This double-peak feature is indeed a beautiful signature of spin fractionalization to Majorana fermions and a gauge field that occurs in two separate temperature regimes, leaving their fingerprint on the $C_v$ curve. In the following, we discuss the thermodynamics of the 3D Kitaev model on the hyperhoneycomb lattice and present our tensor network results. In particular, we show how the  signatures of spin fractionalization can be observed in different physical quantities such as spin correlations, thermal entropy, and bond entanglement entropy. In the very low-temperature regime $T\ll\Delta$, where the TN simulations are challenging, we provide the results for the Chern number and thermal Hall conductivity directly from the single-particle spectrum, in order to offer a complete picture for the thermodynamics of the 3D Kitaev model.    

\subsection{High-temperature spin-ordering crossover}

\begin{figure*}
	\centerline{\includegraphics[width=18cm]{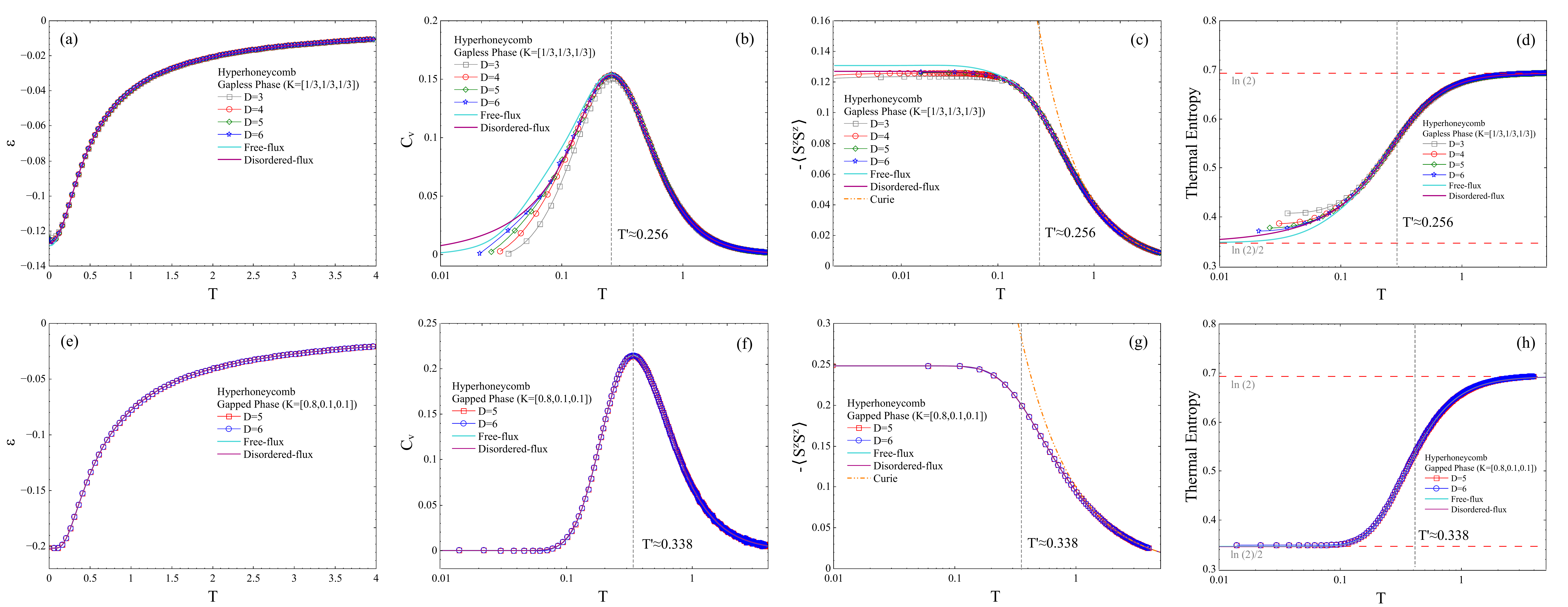}}
	\caption{(Color online) (a) Thermal energy per-site, $\varepsilon$, (b) specific heat $C_v$, (c) nearest-neighbor spin-spin correlation and (d) thermal entropy of the Kitaev model on the hyperhoneycomb lattice at the isotropic point $K_x=K_y=K_y=1/3$ (gapless region of the phase diagram Fig.~\ref{Fig:3Dphasediag}). (e) The energy, (f) specific heat, (g) NN spin-spin correlation and (h) thermal entropy of the Kitaev model for $K_x=0.8, K_z=K_y=0.1$ (gapped region of the phase diagram Fig.~\ref{Fig:3Dphasediag}). See text for details.}
	\label{Fig:Thermodynamics}
\end{figure*}

Let us start our discussion by focussing on the isotropic point in the gapless region of the phase diagram, i.e., ${K_x=K_y=K_z=1/3}$. Approaching from the high-temperature regime where the system is in a spin paramagnet (spin gas) phase, we first hit the second peak in the specific heat, $C_v=\partial \varepsilon_0(T)/\partial T$. For the case of the hyperhoneycomb lattice this is located at the crossover temperature $T'\approx 0.256$ (see Fig.~\ref{Fig:Thermodynamics}-(b)). 

In order to understand the nature of this peak, let us remind that in the parton configuration, the Kitaev model is characterized by Majorana fermions, which are local objects, coupled to a $\Zd$ gauge field (see also Sec.~\ref{sec:model}). In this language, the kinetic energy of the Majorana fermions, $i\expectval{c_ic_j}$, is precisely equivalent to the nearest-neighbor (NN) spin correlation, $S^{\gamma\gamma}(T)\equiv\expectval{S_i^\gamma S_j^\gamma}_T$. We, therefore, calculated the NN spin correlation of the Kitaev model on the hyperhoneycomb lattice as shown in Fig.~\ref{Fig:Thermodynamics}-(c). Interestingly, the high-temperature crossover coincides with the onset of NN spin correlations at ${T'\approx 0.256}$, in the same location as that of the second peak of the specific heat, which is the temperature at which the spin degrees of freedom begin to be locally fractionalized (see also Fig.~\ref{Fig:Double_peak_sketch}). Our TN simulation further shows that the system is a conventional paramagnet in the high-temperature region ${T>T'}$ and the NN spin correlation obeys the Curie-Weiss behavior, $S^{\gamma\gamma}\propto K_\gamma/T$. In the opposite extreme limit, however, the NN spin correlation reaches its ${T=0}$ saturation value, i.e., ${S^{zz}=0.5248/4}$, characteristic of the gapless Kitaev QSL \cite{Baskaran2007}.

This NN spin ordering is a purely local phenomenon, and its features and its location do not strictly depend on things like the lattice size or geometry \cite{Eschmann2020,Nasu2014a,Nasu2015}. It is therefore a thermal crossover which can be revealed by the second peak of the specific heat. In our TN simulations, this thermal crossover at the second peak can be better identified using entanglement scaling. Fig.~\ref{Fig:Tc_Scaling}-(a) demonstrates the scaling of the location of the high-temperature peak $T'$ versus inverse bond dimension $1/D$. One can clearly see that the location of $T'$ is invariant with respect to the change of bond dimension, implying the crossover nature of $ T' $, instead of being a true thermal transition.

The thermal entropy of the system is also given by ${\mathcal{S}_T=\mathcal{S}_\infty-\int_T^{\infty}  C_v(T)\, d \ln T}$, where ${\mathcal{S}_\infty=\ln 2}$ is the maximum entropy density corresponding to an equally weighted, infinite-temperature Gibbs state. As expected, the thermal entropy saturates at ${\mathcal{S}_\infty}$ in the deep paramagnetic phase. However, the system releases exactly half of its entropy at the crossover temperature $T'$ and reaches the plateau $\ln 2/2$. This entropy is related to the contribution of the Majorana fermions to the specific heat of the Kitaev model which is released by moving from high-temperature to the low-temperature regime across the crossover point. The plateau of thermal entropy for the Kitaev model on the hyperhoneycomb lattice is shown in Fig.~\ref{Fig:Thermodynamics}-(d).    

In the corners of the 3D Kitaev phase diagram,
%where one of the couplings dominates the two others,
in the toric code limit,
the system is fully gapped. It is, therefore, reasonable to expect that the crossover to high-temperature paramagnetic spin gas occurs at a relatively larger temperature. The lower panels (e)-(h) of Fig.~\ref{Fig:Thermodynamics} illustrate the energy, specific heat, NN spin correlation, and thermal entropy of the Kitaev model for ${K_x=0.8, K_z=K_y=0.1}$ on the hyperhoneycomb lattice. The crossover temperature for this gapped region is located at ${T'=0.338}$ as shown in the plot of specific heat (Fig.~\ref{Fig:Thermodynamics}-(f)). Similar to the isotropic point, here the NN spin correlation saturates as well to $S^{zz}\approx 0.248$ for $T<T'$ down to zero temperature which is accompanied by release of half of the entropy, implying again the fractionalization of original spins. 

Last but not least, we have benchmarked our TN simulations against the corresponding quantities extracted from the effective tight-binding Hamiltonian~\eqref{eq:H-Majorana}. By assuming static fixed-flux configurations, thermal fluctuations of the gauge fields can be ignored and only Majorana fermions are retained. It is then straightforward to see that in the Majorana basis, the thermal average over spin correlation and energy reads
\bea
S_{\{u_{ij}\}}^{\gamma\gamma}(T)&=&-\frac{i}{2V} \sum_{\langle i,j \rangle \mu} u_{ij}  \psi_\mu(i) \psi^*_\mu (j) \tanh\left(\dfrac{\beta \epsilon_{\mu,\{u_{ij}\}}}{2} \right), \nonumber \\
E_{\{u_{ij}\}}(T)&=&-\frac{1}{4} \sum_\mu \dfrac{\epsilon_{\mu,\{u_{ij}\}}}{2} \tanh\left(\dfrac{\beta \epsilon_{\mu,\{u_{ij}\}}}{2}\right),
\label{eq:E_corr}
\eea
where $ {\psi_\mu(i) } $ is the $ i^\mathrm{th} $ component of the normalized complex eigenvector of 
the Kitaev Hamiltonian, and the sum on $\mu$ runs over half of the single-particle spectrum with non-negative $ {\epsilon_\mu>0} $.
The solid lines in all panels of Fig.~\ref{Fig:Thermodynamics} correspond to the results for the ground-state gauge ansatz, i.e., the free-flux sector, as well as those averaged over random flux configurations. We find remarkable agreements between the disorder-averaged data and the TN results, in particular above the gauge-field-disordering transition. Besides, the results of free-flux and random gauge configurations become indistinguishable above the spin-disordering crossover in the thermal paramagnet phase.

\subsection{Low-temperature gauge-ordering transition}
Next, we discuss the thermodynamics of the Kitaev model on the hyperhoneycomb lattice below the crossover temperature, i.e., $T<T'$. As already pointed out previously, the spin degrees of freedom are fractionalized to the itinerant Majorana fermions and a $\Zd$ gauge field at the crossover temperature $T'$. While the formation of Majorana fermions can be captured perfectly by NN spin correlation (since the fermions are local), there is however no direct local measure to reveal the thermodynamics of gauge fields. The gauge fields in the 3D Kitaev model are defined by non local strings which go around the plaquettes of the hyperhoneycomb lattice. Associating a nonlocal flux operator to each plaquette $p$ of the lattice as
\bea
\widehat{W}_p=\prod_{\expectval{i,j},\gamma\in p} S_i^\gamma S_j^\gamma=\prod_{\expectval{i,j},\gamma\in p} (-i\hat{u}^\gamma_{i,j}),
\eea  
the gauge structure can be captured by the $\pm 1$ eigenvalues of the $\widehat{W}_p$ operators,  which are integrals of motion of the Kitaev Hamiltonian \eqref{eq:H-Kitaev}. The QSL ground state of the Kitaev model at $T=0$ is therefore distinguished by eigenvalues $W_p=1=e^{i0}$ ($-1=e^{i\pi}$), indicating the presence of a $\Zd$ gauge field on the plaquettes with uniform zero ($\pi$) flux. Below the low-temperature phase transition, i.e., $T<T_c$, the system is therefore expected to be in a state with an ordered gauge structure.

By increasing the temperature above the $T>T_c$, thermal fluctuations break the patterns of gauge loops and allow the formation of loops with different shapes and sizes, hence a disordered gauge background with $W_p=0$ is stabilized in the system. The intermediate disordered gauge region continues to persist for $T_c\leq T \leq T'$ until it is totally thermalized above the crossover temperature to the {trivial} paramagnet phase. 

The low-temperature gauge ordering transition is revealed by the first peak in the specific heat of the Kitaev model as sketched in Fig.\ref{Fig:Double_peak_sketch}. Previous Monte Carlo studies \cite{Eschmann2020,Mishchenko2017,Nasu2014,Nasu2014a,Nasu2015} have revealed that the thermal gauge ordering transition is located at very low temperature of the order $10^{-2}K-10^{-3}K$ depending on the lattice size, geometry and spatial dimension. In contrast to the crossover temperature, the low-temperature gauge ordering is an actual thermal phase transition. 

\begin{figure*}  
	\centerline{\includegraphics[width=18cm]{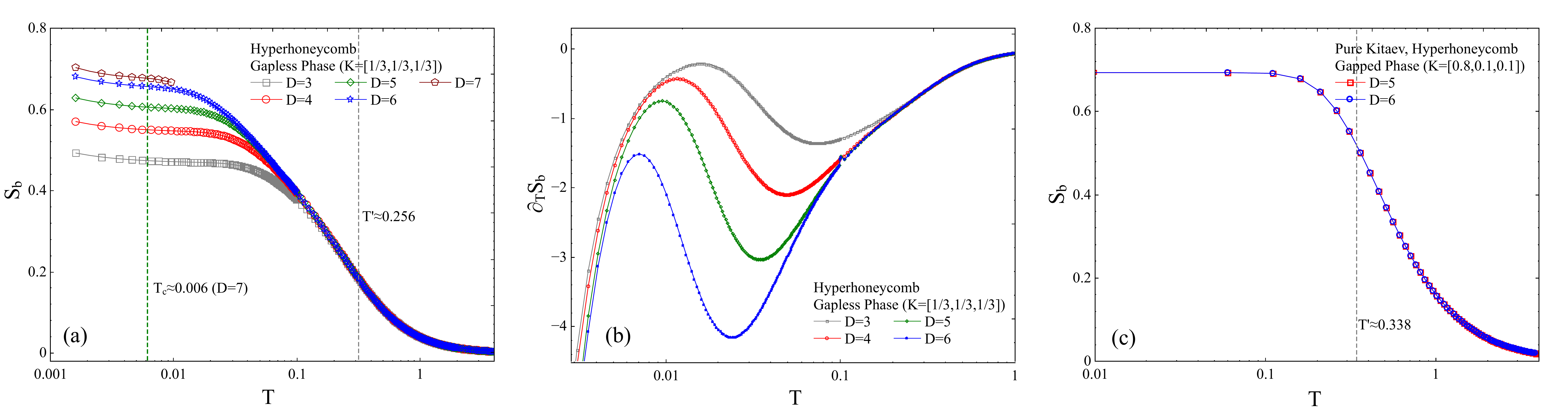}}
	\caption{(Color online) Bond entanglement entropy, $S_{b}$, of the Kitaev model on the hyperhoneycomb lattice at (a) the isotropic point $K_x=K_y=K_y=1/3$ and (c) deep inside the gapped region $K_x=0.8, K_z=K_y=0.1$ for different bond dimension $D$. (b) The first derivative of the $S_{VN}$ of the isotropic point which pinpoints the location of the thermal transition point $T_c$ for different bond dimensions. }
	\label{Fig:Svn}
\end{figure*}

The local simple-update and the mean-field environment that we used in the TgPEPS algorithm does not allow us to accurately simulate very low temperatures of the order $ T\lesssim T_c\sim K/100$ due to poor convergence of the SU in this temperature regime. We where therefore unable to calculate the expectation value of the $\widehat{W}_p$ operators as a direct probe for capturing the gauge ordering. However, as an indirect signature of the entangled gauge loops around the plaquettes, we calculated the \emph{bond entropy}, 
\bea
S_{b}=-\sum_i \lambda_i^2 \log \lambda_i^2,
\eea
for the matrices $\lambda_i$  of singular values at bond $i$ in the TN. These singlar values contain  information about the entanglement and correlation structure on virtual bonds of the lattice. Since every link of the lattice is a part of a closed plaquette, the entangled gauge loops which span along the plaquettes of the lattice leave their fingerprint on bond entanglement. Figure~\ref{Fig:Svn}-(a) show the bond entropy of the Kitaev model on the hyperhoneycomb lattice for different bond dimensions $D$ at the isotropic point. 

\begin{figure} [t!] 
	\centerline{\includegraphics[width=\columnwidth]{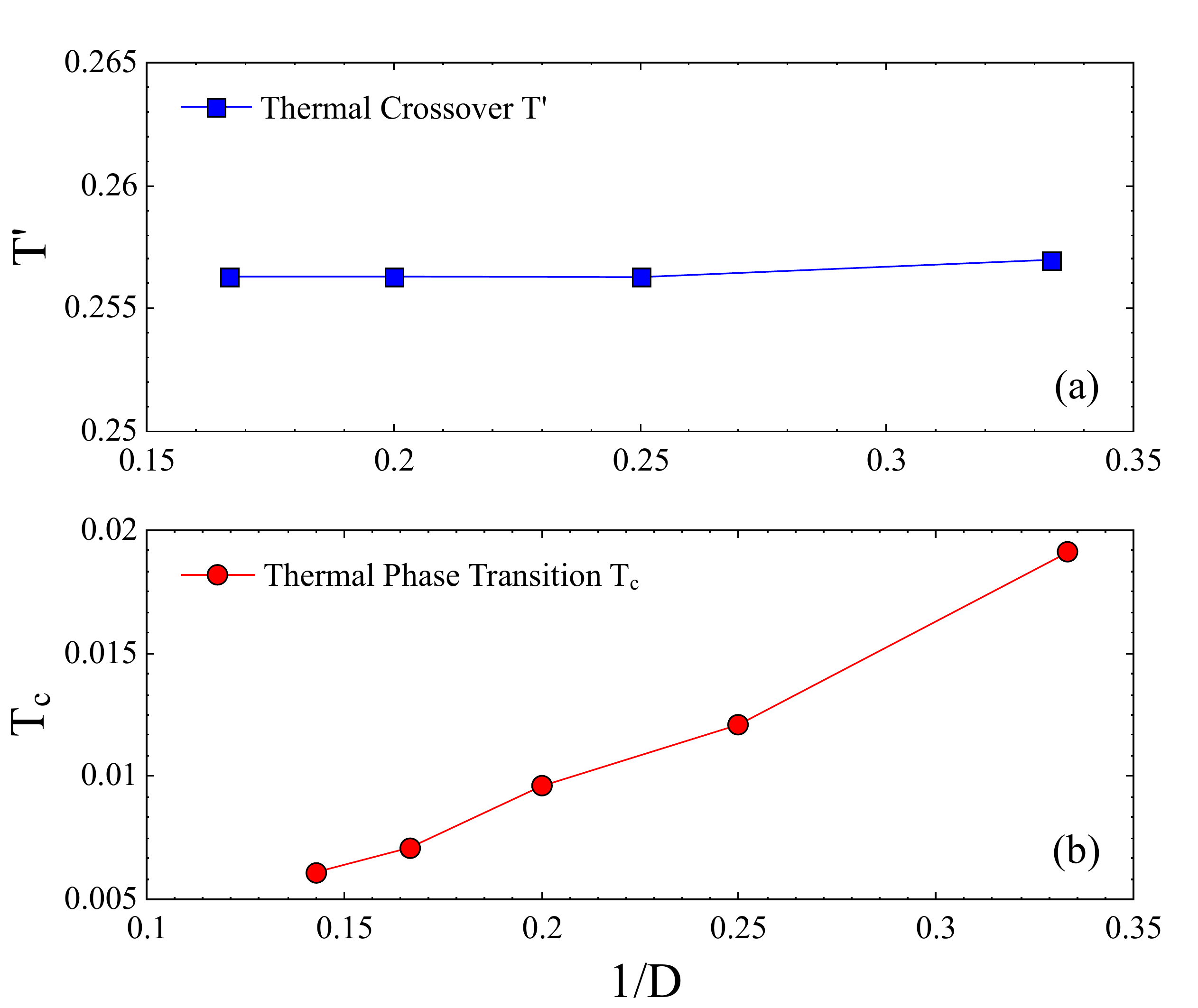}}
	\caption{(Color online) Scaling of the location of the high-temperature crossover $T'$ (upper) and the low-temperature thermal phase transition $T_c$ with inverse bond dimension $1/D$.}	
	\label{Fig:Tc_Scaling}
\end{figure}

The bond entropy is zero in the high-temperature regime in the spin paramagnet phase, as expected from the infinite-temperature limit. As the system is cooled down and the spin fractionalization occurs at the crossover temperature, the bond entropy starts to grow until it reaches the thermal transition point at $T_c$ below which the system is in the Kitaev QSL phase. While our TN specific heat (Fig.\ref{Fig:Thermodynamics}-(b)) is insensitive to the first peak at $T_c$ (due to local SU update and finite bond dimension), it is remarkable that the location of the gauge ordering transition can be captured from the derivative of $S_{b}$ (\ref{Fig:Svn}-(b)). In contrast to the crossover temperature, physical observables, and their derivatives are sensitive to the bond dimension, and are suitable for entanglement scaling (see Figs.~\ref{Fig:Thermodynamics}-(b)-(d) and Figs.~\ref{Fig:Svn}-(a),(b) at the low-temperature regime). Figure \ref{Fig:Tc_Scaling}-(b) shows the scaling of the location of the gauge-ordering thermal transition for the isotropic point for different bond dimensions  $D$. In contrast to $T'$, which is invariant with respect to $D$ (\ref{Fig:Tc_Scaling}-(a)), $T_c$ depends clearly on the bond dimension, which is a typical signature of a diverging correlation length and hence a true quantum phase transition. Our results for the largest considered bond dimension, $D=7$, yield $T_c=0.006$, which is slightly away from the previous QMC estimate $T_c^{QMC}=0.0024$ \cite{Eschmann2020}. However, our simulation also shows a tendency to reach the QMC value by increasing the bond dimension $D$. 

Let us further note that the remaining thermal entropy of the system will be further released at the thermal transition point due to the full ordering of the gauge structure. The entropy plot will then show another drop from the $\ln 2/2$ plateau to $S_T=0$ \cite{Eschmann2020}. Unfortunately, due to the limited convergence of our TN results for $T\lesssim T_c\sim K/100$, we are not able to capture the first entropy release, caused by the thermal fluctuation of visons.

Between the two thermal transitions, there is an intermediate temperature regime that spans over two orders of magnitude, $T_c < T < T'\sim K$ and might, in fact, be the most relevant temperature regime in experimental probes of Kitaev materials. In this regime, one expects to observe the first signatures of fractionalization with the original spins already broken apart into Majorana fermions and a $ \Zd$ gauge field. The latter, however, is still highly disordered in the intermediate regime which prevents the formation of a clean Majorana band structure (as it is the case at strictly zero temperature or, more precisely, the low-temperature transition). Instead one expects to see a disordered Majorana metal (thermal metal), which has already been observed numerically in certain 2D settings \cite{Mishchenko2017,Nasu2014,Nasu2014a,Nasu2015}.

\begin{figure*}  
	\centerline{\includegraphics[width=\linewidth]{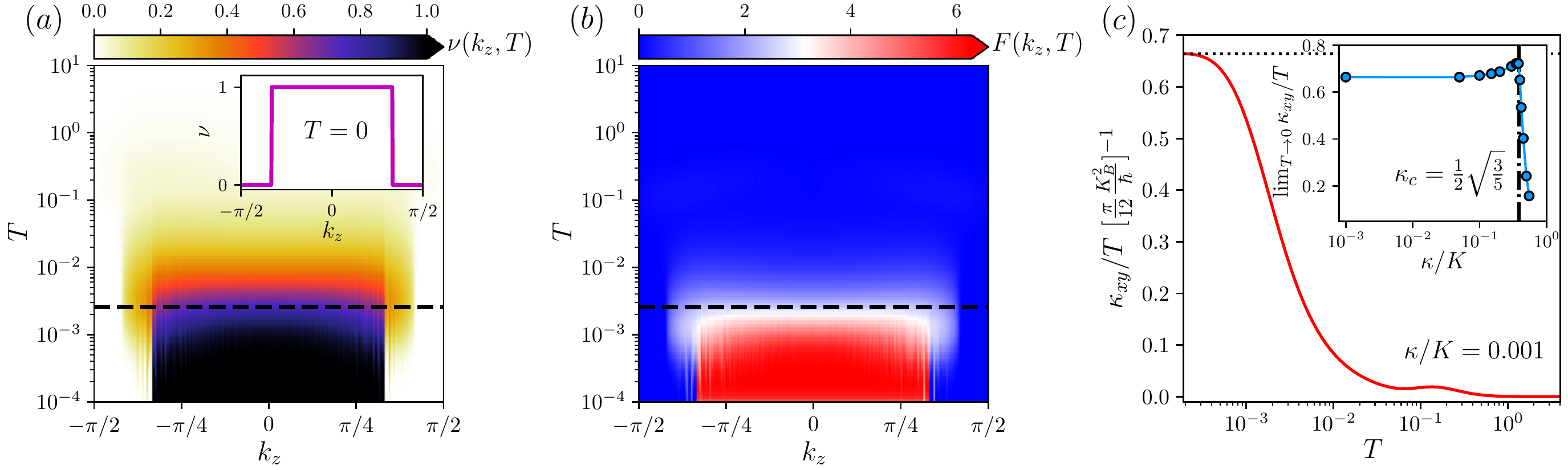}}
	\caption{(Color online) (a) Color plot of Chern number in the plane of $ {(k_z,T)} $ at a small fixed magnetic fields ${\kappa/K=0.001}$. The Chern number is calculated on a single 2D slice of the 3D BZ parameterized by fixed momentum $ k_z $. The inset shows $ k_z $-dependence of the Chern number at zero temperature. (b) Evolution of the momentum-resolved thermal Hall conductivity as function of temperature at fixed ${\kappa/K=0.001}$, showing an abrupt change in the vicinity of gauge-ordering transition $ T_c $ (dashed line) and finally vanishing in the high-temperature region. (c) The thermal Hall conductivity (in the unit of $ \pi/12 $) as a function of temperature.  In the low-temperature regime, the thermal Hall coefficient is finite but non-quantized. The inset displays a sharp drop in the magnitude of the thermal Hall plateau as a function of the field strength, indicating the change in the total number of Weyl points.}
	\label{Fig:chern-Hall-T}
\end{figure*}

While the crossover mechanism at the high-temperature regime is linked to the spin ordering, the low-temperature thermal transition is associated with the fluctuations of gauge fields. As previously pointed out, the ground state of the Kitaev model (at the isotropic point) on different 3D lattice geometries is given in the zero- or $\pi$-flux sector, which are sectors with loop-like objects constructing the boundaries of closed volumes. At the critical point $T_c$, these loops break apart and span the whole lattice. The transition between these two regimes has been argued to be of the second-order type, and in the 3D Ising universality class. Moreover, recent QMC simulations have shown that the thermal phase transition in the gapped regions of the phase diagram of the Kitaev model, e.g., $K\equiv K_x=K_y\ll 1, K_z\to1$, occurs at even lower temperature, of the order $T_c=1.925K_{\rm eff} \ll K/100$ with $K_{\rm eff}\propto\frac{7K^6}{256K_z^5}$ \cite{Nasu2015}, which is beyond the reach of both our TN and exact calculations. 

Here we have calculated the bond entanglement entropy of the gapped phase at $[0.8,0.1,0.1]$ (see Fig.~\ref{Fig:Svn}-(c)). As seen in the figure, this shows a similar behaviour to the isotropic case, indicating an increase of correlations from the paramagnet regime to the gauge disordered region. We expect that at the thermal gauge ordering transition, $S_{b}$ will show another increase due to the QSL ground state at $T=0$.

\subsection{Thermal Hall effect}

Let us now show how the gauge-ordering thermal transition at the isotropic point can be captured precisely by thermal Hall conductivity and Chern number.  As we pointed out in Sec.~\ref{sec:model}, Breaking TRS by a field term opens a gap at the Dirac points of the bulk spectrum of the model and and yields a chiral edge mode with a non-vanishing Chern number, giving rise to a finite thermal Hall conductance when subjected to a thermal gradient. In order to shed light on the relationship between the Majorana band topology and thermal Hall response, we evaluate the Chern number of the effective Majorana Hamiltonian \eqref{eq:H-Kitaev-Field} by slicing the 3D Brillouin zone (BZ) into 2D planes passing through three points 
$ \mathbf{k} = (0, 0, k_z) $, $ \mathbf{k} + \mathbf{q}_2/2 $ and $ \mathbf{k} + \mathbf{q}_3/2 $, where $ \mathbf{q}_2$ and $ \mathbf{q}_3 $ are reciprocal lattice vectors of hyperhoneycomb lattice. This allows to define the total Chern number for any fixed value of $ k_z $, 
\bea
\nu (k_z)= 2\pi \int_{\mathrm{BZ}(k_z)}\frac{d^2 k}{(2\pi)^2} \sum_{\epsilon_{n\mathbf{k}}<0} \Omega^z_{n\mathbf{k}},
\label{eq:Chern}
\eea
where ${\mathbf{\Omega}_{n\mathbf{k}}=i \langle \mathbf{\nabla}_\mathbf{k} u_{n\mathbf{k}}|\times| \mathbf{\nabla}_\mathbf{k} u_{n\mathbf{k}}\rangle} $ is the non-Abelian Berry curvature for the $n^{\mathrm{th}}$ band with energy dispersion $\epsilon_{n\mathbf{k}}$ in the free-flux sector, and $ |u_{nk}\rangle $ denotes the corresponding eigenvector. Here the integral is taken over a single slice of the BZ parameterized by the momentum $ k_z $, and the sum runs over all occupied bands with negative energy. The definition of the Chern number can further be extended to finite temperature as ${ \nu (k_z,T)} $ by replacing $ {\Omega^z_{n\mathbf{k}}\to f(\epsilon_{n\mathbf{k}},T) \Omega^z_{n\mathbf{k}}}  $ in Eq.~\eqref{eq:Chern},  
i.e., $ \Omega^z_{n\mathbf{k}} $ is weighted by the Fermi-Dirac distribution,  ${f(\epsilon,T)=1/(1+\exp(\epsilon/T))}$.

Fig.~\ref{Fig:chern-Hall-T}(a) shows the $ k_z $-dependence of the Chern number as a function of temperature at a fixed magnetic field strength $ {\kappa/K = 0.001} $. Such a small non-zero value of $ \kappa $ allows to capture the low-temperature characteristic features of the system in the absence of magnetic field, for which QMC results are already at hand. In the high temperature region, the system is topologically trivial with ${\nu(k_z)=0} $ due to Berry curvature cancellation of almost equally populated bands. As the system is cooled down to the gauge-ordering temperature $ T_c $, ${ \nu (k_z,T)} $ begins to show characteristic features that are expected to appear at exactly zero temperature: once a plane passes through a gapless Weyl node, located at $ (0, 0, \pm k_{0}) $, the Chern number jumps discontinuously by an amount given by the charge of the Weyl point (see the inset of Fig.~\ref{Fig:chern-Hall-T}(a)).

Next, we discuss the thermal Hall effect, in order to investigate how the changes in topological nature of the system and the presence of Weyl nodes can manifest themselves via nontrivial transport features. The expression for the thermal Hall conductivity of a general non-interacting Majorana Hamiltonian is given by~\cite{Luttinger1964,Qin2011,Matsumoto2011,Go2019} (we set $ {K_B = \hbar = 1} $ for simplicity),
\bea
\kappa_{xy}=-\frac{1}{4\pi T V}\int d\epsilon \epsilon^2 \frac{df}{d\epsilon} \sigma_{xy}(\epsilon),
\label{eq:kappa_THE}
\eea
where $ V $ is the volume of the system, ${f(\epsilon,T)}$ is the Fermi-Dirac distribution, and,
\bea
\sigma_{xy}(\epsilon) &=& 2\pi \sum_{\mathbf{k},\epsilon_{n\mathbf{k}}<\epsilon}\Omega^z_{n\mathbf{k}},
\label{eq:sigma_AHE}
\eea
is the zero-temperature anomalous Hall coefficient for a system with the chemical potential $ \epsilon $. According to Eqs.~\eqref{eq:kappa_THE} and~\eqref{eq:sigma_AHE}, a finite Berry curvature is the essential ingredient for generating the Hall conductance. It is also instructive to look at the momentum-resolved representation of the thermal Hall conductance (MRHC) on each 2D plane of the sliced BZ, to identify the contribution of each plane separately. The MRHC can be defined as $ {\kappa_{xy}(T)= \int dk_z F(k_z,T) } $.  Figs.~\ref{Fig:chern-Hall-T}(b,c) show the $ k_z $-variation of the integrand $ F(k_z,T) $ and the corresponding $ {\kappa_{xy}(T)} $, respectively. One can clearly see that the detailed feature of ${ F (k_z,T)}$ is strongly tied to the behavior of the Chern number ${\nu (k_z,T)} $. Remarkably, the ${ F (k_z,T)}$ exhibits an abrupt change in the vicinity of the transition temperature $ T_c  $, and reaches a finite value in the momentum interval with nontrivial topological features, i.e., $ {|k_z|\leq k_0} $.

Lastly, in Fig.~\ref{Fig:chern-Hall-T}(c), we investigate the behaviour of $ \kappa_{xy} $ in the low-temperature limit, indicating a linear $ T $ dependence with a finite but non-quantized coefficient. To leading order in $ {T\to 0}$, the expression of thermal Hall conductivity in Eq.~\eqref{eq:kappa_THE}  is reduced to 
\bea
\lim_{T\to0}{\kappa_{xy}} =  \frac{\pi T}{12} \sigma_{xy}(0).
\label{eq:low-T-kappa}
\eea
The thermal Hall coefficient of such Weyl spin liquids is proportional to the anomalous Hall coefficient at the Fermi energy, $  {\sigma_{xy}(0) = \int \frac{dk_z}{\textcolor{red}{2} \pi} \nu (k_z)} $, which is proportional to the distance between the Weyl nodes (or equivalently the length of surface Fermi arcs), and hence is not generally an integer topological invariant. The appearance of such non-quantized plateau is reminiscent of Weyl superconductors~\cite{Meng2012}, e.g., in engineered heterostructures with alternating conventional (s-wave) superconductor and topological insulator layers. This behavior can also occur \textit{spontaneously} (with no applied magnetic field) in some variants of 3D chiral superconductors~\cite{Yoshioka2018}, and 2D field-driven $ \mathrm{U}(1) $ spin liquids with Dzyaloshinskii-Moriya interactions~\cite{Gao2019}.

It is worth noting that both the precise location and the total number of Weyl point depend on the strength of the magnetic field, which in turn can leave its fingerprint in the magnitude of low-temperature thermal Hall plateau. This can be best perceived from the inset of Fig.~\ref{Fig:chern-Hall-T}(c), where the value of thermal Hall coefficient (in units of $ \pi/12 $) is shown as a function of magnetic field $ \kappa $. Upon increasing $ {\kappa<1} $, the value of thermal Hall plateau remains fairly constant up to  $ {\kappa_c=\frac{1}{2}\sqrt{\frac{3}{5}}}$~\cite{Hermanns2015}, at which it drastically drops due to the change in the total number of Weyl points from two to six (each pair of Weyl points splits into three, conserving zero net chirality).

\begin{figure*}
	\centerline{\includegraphics[width=18cm]{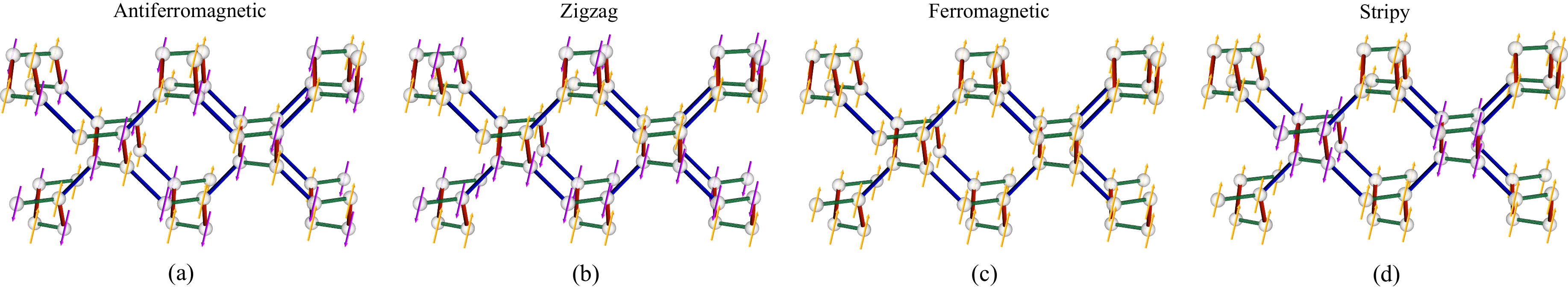}}
	\caption{(Color online) (a) Sketch of the magnetically ordered phases emerging in the vicinity of the QSL regions in the phase diagram of the 3D Kitaev model on the hyperoctagon lattice. (a) The antiferromagnetic, (b) zigzag, (c) ferromagnetic and (d) stripy phases can be identified from the arrows that represent the spin orientation at the vertices.}
	\label{Fig:Ordering}
\end{figure*}

\begin{figure*}  
	\centerline{\includegraphics[width=18cm]{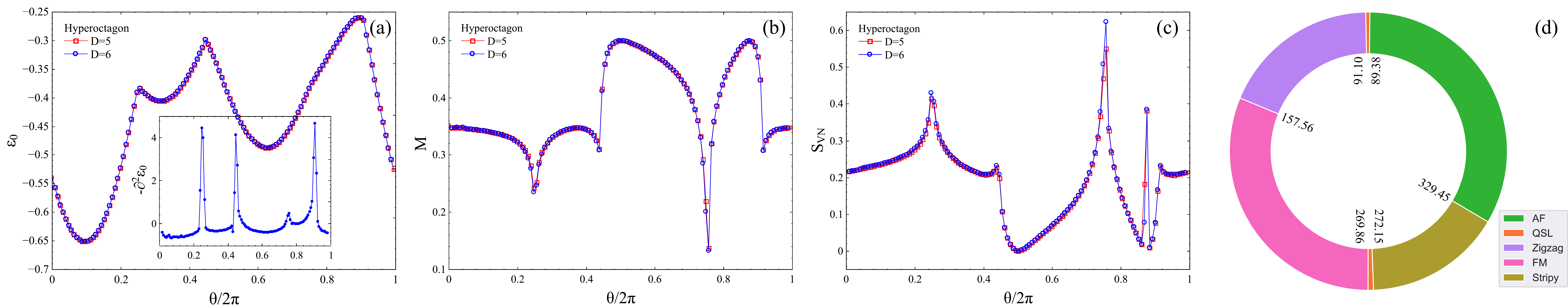}}
	\caption{(Color online) Phase diagram of the Kitaev-Heisenberg model on the hyperoctagon lattice. The phase boundaries can be identified with different quantities as shown in (a) energy per-site $\varepsilon_0$, (b) total magnetization and (c) bond entropy. (d) Phase boundaries in the full parameter space $\theta=[0,2\pi]$.}
		\label{Fig:KH_Phase_Diag}
\end{figure*}

\section{Hyperoctagon Kitaev-Heisenberg Model} 
\label{sec:kitaev_heisenberg}

It has already been argued that the interplay between the crystal-field and spin-orbit coupling in iridates not only results in the emergence of bond-anisotropic Kitaev interactions but also induces isotropic exchange interactions of Heisenberg type \cite{Trebst2017,Chaloupka2010,Chaloupka2013}. It is therefore of particular interest to study the interplay between the Kitaev interaction and Heisenberg exchange coupling. Let us further note that the applicability of QMC simulations is restricted to those 3D Kitaev models whose parton description does not exhibit a sign problem, being this a feature that generally gets lost due to the Heisenberg interaction. This motivates us to analyze the ground state and finite-temprature properties of the Kitaev-Heisenberg (KH) model on 3D lattices in the thermodynamic limit. The Hamiltonian of the KH model is given by    
\be
\label{eq:H-KH}
H_{{\rm KH}}=2K \sum_{\gamma-link} S_i^\gamma S_j^\gamma+J \sum_{\langle i j \rangle} \mathbf{S}_i \cdot \mathbf{S}_j,
\ee
where ${K=\sin\theta}$ and ${J=\cos\theta}$ are the Kitaev and Heisenberg exchange couplings, respectively. 

In the following we consider the hyperoctagon lattice. We first elaborate on the zero-temperature phase diagram of the model, and discuss different magnetically ordered phases hosted in the vicinity of the QSL region in the full parameter space of Hamiltonian \eqref{eq:H-KH}, i.e, $\theta\in[0,2\pi]$. Additionally, we provide further insight into the finite temperature phase transition in the magnetically ordered phase. 

\subsection{$T=0$ Phase Diagram}

Lets us start by investigating the limiting cases of the KH Hamiltonian \eqref{eq:H-KH}. In the extreme regime where the Heisenberg interaction is switched off, i.e., ${\theta=\frac{\pi}{2}}$ (${\theta=\frac{3\pi}{2}}$), the Hamiltonian \eqref{eq:H-KH} reduces to the antiferromagnetic (AFM) (ferromagnetic (FM)) pure Kitaev model and the ground state of the system in these regimes is given by a $\Zd$ QSL phase. In contrast, the opposite limit where $\theta=0$ (${\theta=\pi}$) the system host a trivial antiferromagnetic (ferromagnetic) phase. Additionally, it has been shown that there is a duality (so-called Klein duality~\cite{Chaloupka2010,Kimchi2014}) between specific points (angles) of the phase diagram of the Kitaev-Heisenberg models: for specific angles $\theta$ there is a transformation to another angle $\tilde{\theta}$ such that $\tan\tilde{\theta}=-\tan\theta-1$. The mapping immediately reveals that by transforming the exact FM angle at $\theta=\pi$, another hidden SU(2)-symmetric point is revealed at $\theta=-\frac{\pi}{4}$ for which the Kitaev term vanishes. Solving the resulting Heisenberg Hamiltonian at $\theta=-\frac{\pi}{4}$ yields a FM ground state which, after transforming back to the original spins, maps to a stripy state. Thanks to this duality, we find that the AFM point ($\theta=0$) is also isomorphic to $\theta=\frac{3\pi}{4}$, which after the same transformation and rotations leads to the zigzag state.

A sketch of the magnetically ordered phases on the hyperoctagon lattice is shown in Fig.~\ref{Fig:Ordering}. One can empirically check the Klein duality by considering a $4$-sublattice transformation between the FM and stripy phases as well as the AFM and the zigzag states. Let us further note that this transformation holds true for both 2D and 3D bipartite lattices, with the exception of those systems in which the AFM phase is frustrated.    

\begin{figure*}  
	\centerline{\includegraphics[width=18cm]{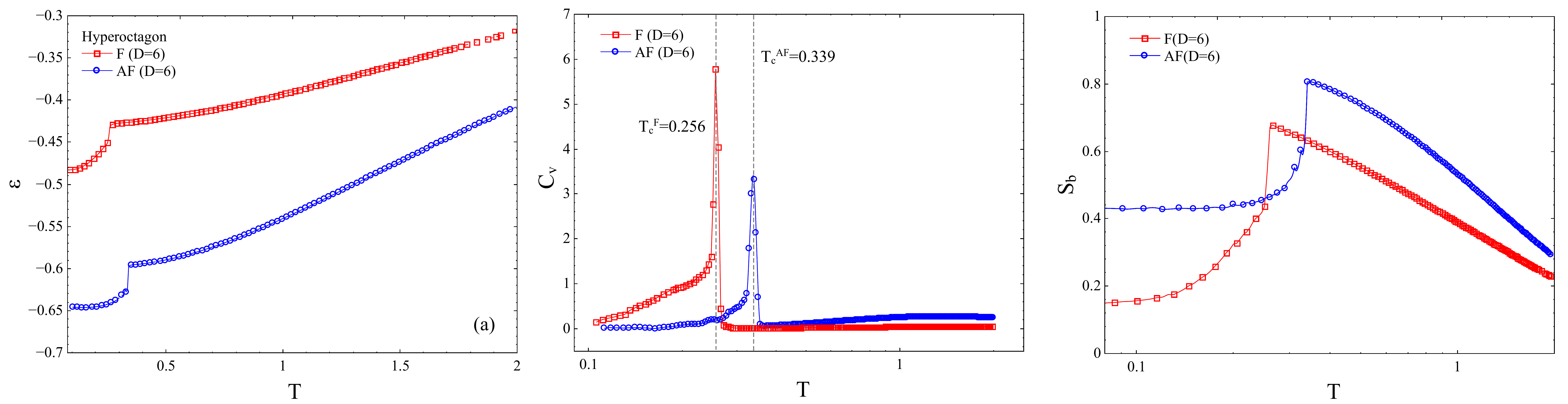}}
	\caption{(Color online) (a) Energy, (b) specific heat $C_v$ and (c) bond entropy, $S_{b}$, of the KH model on the hyperoctagon lattice versus $T$, deep inside the antiferromagnetic ($\theta=34^o$) and ferromagnetic ($\theta=217^o$) phases.}
	\label{Fig:HO_F_AF_FinT}
\end{figure*}

Away from these points, the KH model is no longer exactly solvable. We therefore resort to TN simulations in such regimes. Let us further stress that the KH model is not tractable by QMC, except at the pure Kitaev limit $\theta=\frac{\pi}{2},\frac{3\pi}{2}$. Even in this limit, the exact analytical solution is not available due to the inapplicability of Lieb's theorem on the hyperoctagon lattice~\cite{Hermanns2014,Mishchenko2017,OBrien2016}. However, the $32$-site unit cell used in the thermodynamic-limit gPEPS/TgPEPS methods allows for the Klein duality and is thus expected to capture all the symmetry-breaking patterns in both zero and finite-T phase diagrams.

Fig.~\ref{Fig:KH_Phase_Diag}-(d) illustrates the full TN phase diagram at zero temperature of the KH model on the hyperoctagon lattice. The phase boundaries can be obtained based on a number of different signatures such as the ground state energy and its derivatives, magnetization and bond entropy, as shown in Fig.~\ref{Fig:KH_Phase_Diag}-(a-c). We find signatures of the AFM order for $\theta_{\mathrm{AF}}=[-30^\circ,89.38^\circ]$ and the stripy order for $\theta_{\mathrm{stripy}}=[91.10^\circ,157.56^\circ]$. The FM and zigzag orders are also stabilized for $\theta_{\mathrm{FM}}=[157.56^\circ,269.86^\circ]$ and $\theta_{\mathrm{zigzag}}=[272.15^\circ,329.45^\circ]$, respectively.

Let us further note that the two QSL phases for the AFM and FM Kitaev exchange couplings emerge for $\theta_{\mathrm{AF-QSL}}=[89.38^\circ,91.10^\circ]$ and $\theta_{\mathrm{FM-QSL}}=[269.86^\circ,272.15^\circ]$ regions, respectively in the phase diagram of the KH model. The phase boundaries of the transition between the QSL phases and the ordered phase as well as the transition between the magnetically ordered phases seems to be of first-order, since it shows a discontinuity in the first-order derivative of the ground state energy (also visible with sharp discontinuous peaks in the second derivative of energy as depicted in the inset of Fig.~\ref{Fig:KH_Phase_Diag}-(a)). These findings are in agreement with previous mean-field studies on other 3D Kitaev structures such as the hyperhoneycomb lattice \cite{Lee2014}. Besides, the sharp discontinuity at other quantities such as total magnetization (Fig.~\ref{Fig:KH_Phase_Diag}-(b)) and $S_{b}$ (Fig.~\ref{Fig:KH_Phase_Diag}-(c)) is typical of first-order transitions. 

The two spin liquid phases in the phase diagram are best identified by a vanishing total magnetization as shown in Fig.~\ref{Fig:KH_Phase_Diag}-(b) for $\theta=[89.38^\circ,91.10^\circ]$ and $\theta=[269.86^\circ,272.15^\circ]$. While the zero magnetization is not fully visible in Fig.~\ref{Fig:KH_Phase_Diag}-(b), we examined the QSL region with our TN simulations with more resolution (tiny steps between the couplings) and confirmed the location of the QSL phase boundaries with the gPEPS technique (not shown here). See a similar simulation for the hyperhoneycomb lattice in Ref.~\cite{Jahromi2019}.

\subsection{Magnetically ordered phases at finite-$T$}

In Sec.~\ref{sec:zero_T_pure} and \ref{sec:fin_T_pure}, we investigated the QSL phase of the pure Kitaev model for the AFM exchange couplings. In the absence of external field, the FM QSL ground state has also been shown to have the same thermodynamic properties as that of the AFM case \cite{Hickey2019}. Here we further elaborate on the thermodynamics of the ordered phases in the phase diagram of the KH model on the hyperoctagon lattice. More specifically, we will focus on the thermal phase transition in the FM and AFM ordered regions of the phase diagram in the thermodynamic limit, keeping in mind that the stripy and zigzag phases respectively are related to the FM and AFM states according to the Klein duality.

In contrast to the QSL phases, in which the double-peak nature of the specific heat is associated with the two species of elementary excitations (namely, itinerant Majoranas and visons) with different energy scales, the magnetically ordered phases of the KH model are trivial, showing no signature of fractionalization. We, therefore, expect to observe only one peak in the specific heat of the KH model in the finite-temperature regime of the ordered regions.

In what follows, we fix $\theta=34^\circ$ and $\theta=217^\circ$ deep in the AFM and FM phases,  respectively, and use the temperature $T$ as tuning parameter. Figure.~\ref{Fig:HO_F_AF_FinT}-(a-c) shows the energy, specific heat, and the bond entropy in both the FM and AFM regions of the KH model on the hyperoctagon lattice obtained with the TgPEPS method. As expected, we observe only a single transition between the zero-temperature spin-ordered and high-temperature spin-disordered phases. This transition happens at $T^{\mathrm{AF}}_c=0.51$ and $T^{\mathrm{FM}}_c=0.53$ for the AFM and FM cases, respectively. Cooling down the system from the high-temperature regime  spontaneously breaks the spin inversion symmetry down at $T_c$ for both FM and AFM phases, indicating a thermal phase transition between the high- and low-temperature regimes. Besides, the $T_c$ in both FM and AFM phases is of the same order of magnitude, indicating a similar robustness in the presence of thermal fluctuations.

\section{Conclusions and discussion} 
\label{sec:conclude}

The Kitaev model is one of the first examples of quantum spin liquids with fascinating properties such as topological order and fractional excitations. While the 2D version on the honeycomb lattice has been shown to host a non-abelian topologically ground state at zero temperature \cite{Kitaev2006}, the 3D version is known to have distinct properties in the nodal manifold such as topologically protected Weyl nodes \cite{OBrien2016}. Most importantly, it has been suggested that iridate compounds such as $\beta-\rm{Li_2IrO_3}$ and $\gamma-\rm{Li_2IrO_3}$ are relevant platforms for experimental realizations of 3D Kitaev materials \cite{Takayama2015,Modic2014}. On top of that, the Kitaev model is a fascinating playground for a deeper understanding of the fractionalization mechanism which occurs in QSLs when the system is cooled down to zero temperature. It is the perfect theory lab to study quantum phases of matter. 

While the 2D Kitaev model has been widely studied both analytically and numerically, its 3D version is less explored due to the lack of efficient numerical techniques capable of simulating 3D structures. Current state of the art QMC techniques can only simulate the pure Kitaev model in specific gauge sectors, where the system remains sign-free, and are inapplicable in the presence of relevant perturbations such as Heisenberg interactions and magnetic fields. Other techniques such as exact diagonalization or mean-field approximation are also not free from limitations, such as large finite-size effects and/or deficiency in capturing thermal properties of experimental relevance. 

In this paper we proposed an advanced and efficient tensor network algorithm based on graph-projected entangled-pair state for simulating both the ground state and thermodynamic properties of  3D Kiatev qantum spin liquids. In order to demonstrate the accuracy and efficiency of our algorithms, we simulated the 3D Kitaev model on the hyperhoneycomb lattice and mapped out the full phase diagram of the system at zero temperature. We further simulated its thermal density matrix  and investigated the thermodynamic properties of both gapless and gapped regions in the phase diagram. We showed how the original spin degrees of freedom are fractionalized to itinerant-Majorana fermions and a static gauge field, leaving their fingerprints on quantities such as specific heat, spin correlations, thermal entropy, and bond entropy. In particular, we found that by cooling down the system from the high-temperature regime, where the system is in a spin paramagnet phase, we first hit a crossover temperature at $T'\sim K$ ($K$ being the Kitaev exchange coupling), below which the spins are fractionalized to a spin-ordered phase equivalent to itinerant-Majorana fermions and a disordered gauge field. While the spins remain ordered down to $T=0$, the gauge field remains disordered in an intermediate regime $T_c < T < T'\sim K$ due to thermal fluctuations, which prevent  the gauge fields (loop structures) from becoming ordered. Eventually, for $T\lesssim T_c\sim K/100$ the gauge structure becomes ordered as well, and the system ends up being in a loop soup, i.e, the Kitaev QSL. 

Let us note that the spin-ordering thermal crossover at $T'$ and the thermal gauge ordering transition at $T_c$ can be identified with two separate peaks in the specific heat and each is accompanied by releasing half of the thermal entropy of the system. While the crossover temperature $T'$ has been shown to be captured readily both numerically and experimentally, the thermal transition point $T_c$ is beyond the reach of both experiments and our TN simulation. However, we showed convincing proof that our TN simulations work perfectly in the intermediate to high-temperature regimes which is also the most relevant regime for the experiment, as well. The poor convergence of our thermal TN algorithm in the low-temperature regime is typical of all thermal TN algorithm based on purification and is related to the large entanglement growth during real-time evolution close to the critical point and in the ground state of QSL phases \cite{Czarnik2019,Bruognolo2017,Chen2020}. 

In order to provide a complete picture about the thermodynamic properties of the Kitaev spin liquid particularly in the low-temperature regime, we calculated the Chern number and the thermal Hall conductivity. We showed that the low-temperature gauge-ordering transition at $T_c$ is associated with the non-trivial Majorana band topology, yielding a chiral edge mode that has a non-vanishing Chern number for $T<T_c$. We further saw that the changes in the topological nature of the system below $T_c$ is related to the presence of Weyl nodes, which is manifested via nontrivial transport features, i.e., a finite thermal Hall conductivity that is non-zero in the QSL phase ($T<T_c$) and vanishes above the low-temperature gauge-ordering thermal transition. 

Away from the pure Kitaev point, we studied the phase diagram of the 3D Kitaev-Heisenberg model on the hyperoctagon lattice  at zero and  finite-temperatures. and captured the magnetically ordered phases in the vicinity of the QSL regions. Our simulations also confirme that, in contrast to the QSL phase which has a double-peak feature in the specific heat, the single thermal phase transition in magnetically ordered phases occurs in the context of Landau symmetry-breaking. 

The tensor network techniques introduced in this study can be used as efficient tools for a deeper understanding of the thermodynamics of strongly correlated systems on any dimension and lattice geometry. We, therefore, believe that they are of potential interest for benchmarking both numerical and experimental studies, and will become essential for the discovery of new phase of matter. 

\section*{Acknowledgements}
The CPU time from ATLAS HPC cluster at DIPC is acknowledged. Support from Ikerbasque is also acknowledged. 

\bibliography{references}

\begin{thebibliography}{90}%
\makeatletter
\providecommand \@ifxundefined [1]{%
 \@ifx{#1\undefined}
}%
\providecommand \@ifnum [1]{%
 \ifnum #1\expandafter \@firstoftwo
 \else \expandafter \@secondoftwo
 \fi
}%
\providecommand \@ifx [1]{%
 \ifx #1\expandafter \@firstoftwo
 \else \expandafter \@secondoftwo
 \fi
}%
\providecommand \natexlab [1]{#1}%
\providecommand \enquote  [1]{``#1''}%
\providecommand \bibnamefont  [1]{#1}%
\providecommand \bibfnamefont [1]{#1}%
\providecommand \citenamefont [1]{#1}%
\providecommand \href@noop [0]{\@secondoftwo}%
\providecommand \href [0]{\begingroup \@sanitize@url \@href}%
\providecommand \@href[1]{\@@startlink{#1}\@@href}%
\providecommand \@@href[1]{\endgroup#1\@@endlink}%
\providecommand \@sanitize@url [0]{\catcode `\\12\catcode `\$12\catcode
  `\&12\catcode `\#12\catcode `\^12\catcode `\_12\catcode `\%12\relax}%
\providecommand \@@startlink[1]{}%
\providecommand \@@endlink[0]{}%
\providecommand \url  [0]{\begingroup\@sanitize@url \@url }%
\providecommand \@url [1]{\endgroup\@href {#1}{\urlprefix }}%
\providecommand \urlprefix  [0]{URL }%
\providecommand \Eprint [0]{\href }%
\providecommand \doibase [0]{http://dx.doi.org/}%
\providecommand \selectlanguage [0]{\@gobble}%
\providecommand \bibinfo  [0]{\@secondoftwo}%
\providecommand \bibfield  [0]{\@secondoftwo}%
\providecommand \translation [1]{[#1]}%
\providecommand \BibitemOpen [0]{}%
\providecommand \bibitemStop [0]{}%
\providecommand \bibitemNoStop [0]{.\EOS\space}%
\providecommand \EOS [0]{\spacefactor3000\relax}%
\providecommand \BibitemShut  [1]{\csname bibitem#1\endcsname}%
\let\auto@bib@innerbib\@empty
%</preamble>
\bibitem [{\citenamefont {Savary}\ and\ \citenamefont
  {Balents}(2017)}]{Savary2017}%
  \BibitemOpen
  \bibfield  {author} {\bibinfo {author} {\bibfnamefont {L.}~\bibnamefont
  {Savary}}\ and\ \bibinfo {author} {\bibfnamefont {L.}~\bibnamefont
  {Balents}},\ }\href {\doibase 10.1088/0034-4885/80/1/016502} {\bibfield
  {journal} {\bibinfo  {journal} {Reports on Progress in Physics}\ }\textbf
  {\bibinfo {volume} {80}},\ \bibinfo {pages} {016502} (\bibinfo {year}
  {2017})}\BibitemShut {NoStop}%
\bibitem [{\citenamefont {Levin}\ and\ \citenamefont {Wen}(2006)}]{Levin2006}%
  \BibitemOpen
  \bibfield  {author} {\bibinfo {author} {\bibfnamefont {M.}~\bibnamefont
  {Levin}}\ and\ \bibinfo {author} {\bibfnamefont {X.~G.}\ \bibnamefont
  {Wen}},\ }\href {\doibase 10.1103/PhysRevLett.96.110405} {\bibfield
  {journal} {\bibinfo  {journal} {Physical Review Letters}\ }\textbf {\bibinfo
  {volume} {96}},\ \bibinfo {pages} {110405} (\bibinfo {year} {2006})},\
  \Eprint {http://arxiv.org/abs/0510613} {arXiv:0510613 [cond-mat]}
  \BibitemShut {NoStop}%
\bibitem [{\citenamefont {Kitaev}(2006)}]{Kitaev2006}%
  \BibitemOpen
  \bibfield  {author} {\bibinfo {author} {\bibfnamefont {A.}~\bibnamefont
  {Kitaev}},\ }\href {\doibase 10.1016/j.aop.2005.10.005} {\bibfield  {journal}
  {\bibinfo  {journal} {Annals of Physics}\ }\textbf {\bibinfo {volume}
  {321}},\ \bibinfo {pages} {2} (\bibinfo {year} {2006})},\ \Eprint
  {http://arxiv.org/abs/0506438} {arXiv:0506438 [cond-mat]} \BibitemShut
  {NoStop}%
\bibitem [{\citenamefont {Jahromi}\ and\ \citenamefont
  {Langari}(2017)}]{Jahromi2017}%
  \BibitemOpen
  \bibfield  {author} {\bibinfo {author} {\bibfnamefont {S.~S.}\ \bibnamefont
  {Jahromi}}\ and\ \bibinfo {author} {\bibfnamefont {A.}~\bibnamefont
  {Langari}},\ }\href {\doibase 10.1088/1751-8121/aa5db6} {\bibfield  {journal}
  {\bibinfo  {journal} {Journal of Physics A: Mathematical and Theoretical}\
  }\textbf {\bibinfo {volume} {50}},\ \bibinfo {pages} {145305} (\bibinfo
  {year} {2017})},\ \Eprint {http://arxiv.org/abs/1512.00756}
  {arXiv:1512.00756} \BibitemShut {NoStop}%
\bibitem [{\citenamefont {Kitaev}(2003)}]{Kitaev2003}%
  \BibitemOpen
  \bibfield  {author} {\bibinfo {author} {\bibfnamefont {A.~Y.}\ \bibnamefont
  {Kitaev}},\ }\href {\doibase 10.1016/S0003-4916(02)00018-0} {\bibfield
  {journal} {\bibinfo  {journal} {Annals of Physics}\ }\textbf {\bibinfo
  {volume} {303}},\ \bibinfo {pages} {2} (\bibinfo {year} {2003})},\ \Eprint
  {http://arxiv.org/abs/9707021} {arXiv:9707021 [quant-ph]} \BibitemShut
  {NoStop}%
\bibitem [{\citenamefont {Levin}\ and\ \citenamefont {Wen}(2005)}]{Levin2005}%
  \BibitemOpen
  \bibfield  {author} {\bibinfo {author} {\bibfnamefont {M.~A.}\ \bibnamefont
  {Levin}}\ and\ \bibinfo {author} {\bibfnamefont {X.~G.}\ \bibnamefont
  {Wen}},\ }\href {\doibase 10.1103/PhysRevB.71.045110} {\bibfield  {journal}
  {\bibinfo  {journal} {Physical Review B - Condensed Matter and Materials
  Physics}\ }\textbf {\bibinfo {volume} {71}},\ \bibinfo {pages} {045110}
  (\bibinfo {year} {2005})},\ \Eprint {http://arxiv.org/abs/0404617}
  {arXiv:0404617 [cond-mat]} \BibitemShut {NoStop}%
\bibitem [{\citenamefont {Balents}(2010)}]{Balents2010}%
  \BibitemOpen
  \bibfield  {author} {\bibinfo {author} {\bibfnamefont {L.}~\bibnamefont
  {Balents}},\ }\href {\doibase 10.1038/nature08917} {\bibfield  {journal}
  {\bibinfo  {journal} {Nature}\ }\textbf {\bibinfo {volume} {464}},\ \bibinfo
  {pages} {199} (\bibinfo {year} {2010})},\ \Eprint
  {http://arxiv.org/abs/9904169} {arXiv:9904169 [cond-mat]} \BibitemShut
  {NoStop}%
\bibitem [{\citenamefont {Liao}\ \emph {et~al.}(2017)\citenamefont {Liao},
  \citenamefont {Xie}, \citenamefont {Chen}, \citenamefont {Liu}, \citenamefont
  {Xie}, \citenamefont {Huang}, \citenamefont {Normand},\ and\ \citenamefont
  {Xiang}}]{Liao2017}%
  \BibitemOpen
  \bibfield  {author} {\bibinfo {author} {\bibfnamefont {H.~J.}\ \bibnamefont
  {Liao}}, \bibinfo {author} {\bibfnamefont {Z.~Y.}\ \bibnamefont {Xie}},
  \bibinfo {author} {\bibfnamefont {J.}~\bibnamefont {Chen}}, \bibinfo {author}
  {\bibfnamefont {Z.~Y.}\ \bibnamefont {Liu}}, \bibinfo {author} {\bibfnamefont
  {H.~D.}\ \bibnamefont {Xie}}, \bibinfo {author} {\bibfnamefont {R.~Z.}\
  \bibnamefont {Huang}}, \bibinfo {author} {\bibfnamefont {B.}~\bibnamefont
  {Normand}}, \ and\ \bibinfo {author} {\bibfnamefont {T.}~\bibnamefont
  {Xiang}},\ }\href {\doibase 10.1103/PhysRevLett.118.137202} {\bibfield
  {journal} {\bibinfo  {journal} {Physical Review Letters}\ }\textbf {\bibinfo
  {volume} {118}},\ \bibinfo {pages} {137202} (\bibinfo {year} {2017})},\
  \Eprint {http://arxiv.org/abs/1610.04727} {arXiv:1610.04727} \BibitemShut
  {NoStop}%
\bibitem [{\citenamefont {Poilblanc}\ \emph {et~al.}(2019)\citenamefont
  {Poilblanc}, \citenamefont {Mambrini},\ and\ \citenamefont
  {Capponi}}]{Poilblanc2019}%
  \BibitemOpen
  \bibfield  {author} {\bibinfo {author} {\bibfnamefont {D.}~\bibnamefont
  {Poilblanc}}, \bibinfo {author} {\bibfnamefont {M.}~\bibnamefont {Mambrini}},
  \ and\ \bibinfo {author} {\bibfnamefont {S.}~\bibnamefont {Capponi}},\ }\href
  {\doibase 10.21468/scipostphys.7.4.041} {\bibfield  {journal} {\bibinfo
  {journal} {SciPost Physics}\ }\textbf {\bibinfo {volume} {7}} (\bibinfo
  {year} {2019}),\ 10.21468/scipostphys.7.4.041},\ \Eprint
  {http://arxiv.org/abs/1907.03678} {arXiv:1907.03678} \BibitemShut {NoStop}%
\bibitem [{\citenamefont {Jahromi}\ \emph {et~al.}(2018)\citenamefont
  {Jahromi}, \citenamefont {Or{\'{u}}s}, \citenamefont {Kargarian},\ and\
  \citenamefont {Langari}}]{Jahromi2018a}%
  \BibitemOpen
  \bibfield  {author} {\bibinfo {author} {\bibfnamefont {S.~S.}\ \bibnamefont
  {Jahromi}}, \bibinfo {author} {\bibfnamefont {R.}~\bibnamefont {Or{\'{u}}s}},
  \bibinfo {author} {\bibfnamefont {M.}~\bibnamefont {Kargarian}}, \ and\
  \bibinfo {author} {\bibfnamefont {A.}~\bibnamefont {Langari}},\ }\href
  {\doibase 10.1103/PhysRevB.97.115161} {\bibfield  {journal} {\bibinfo
  {journal} {Physical Review B}\ }\textbf {\bibinfo {volume} {97}},\ \bibinfo
  {pages} {115161} (\bibinfo {year} {2018})},\ \Eprint
  {http://arxiv.org/abs/1711.04798} {arXiv:1711.04798} \BibitemShut {NoStop}%
\bibitem [{\citenamefont {ANDERSON}(1987)}]{Anderson1987}%
  \BibitemOpen
  \bibfield  {author} {\bibinfo {author} {\bibfnamefont {P.~W.}\ \bibnamefont
  {ANDERSON}},\ }\href {\doibase 10.1126/science.235.4793.1196} {\bibfield
  {journal} {\bibinfo  {journal} {Science}\ }\textbf {\bibinfo {volume}
  {235}},\ \bibinfo {pages} {1196} (\bibinfo {year} {1987})}\BibitemShut
  {NoStop}%
\bibitem [{\citenamefont {Poilblanc}\ \emph {et~al.}(2014)\citenamefont
  {Poilblanc}, \citenamefont {Corboz}, \citenamefont {Schuch},\ and\
  \citenamefont {Cirac}}]{Poilblanc2014}%
  \BibitemOpen
  \bibfield  {author} {\bibinfo {author} {\bibfnamefont {D.}~\bibnamefont
  {Poilblanc}}, \bibinfo {author} {\bibfnamefont {P.}~\bibnamefont {Corboz}},
  \bibinfo {author} {\bibfnamefont {N.}~\bibnamefont {Schuch}}, \ and\ \bibinfo
  {author} {\bibfnamefont {J.~I.}\ \bibnamefont {Cirac}},\ }\href {\doibase
  10.1103/PhysRevB.89.241106} {\bibfield  {journal} {\bibinfo  {journal}
  {Physical Review B - Condensed Matter and Materials Physics}\ }\textbf
  {\bibinfo {volume} {89}} (\bibinfo {year} {2014}),\
  10.1103/PhysRevB.89.241106},\ \Eprint {http://arxiv.org/abs/1404.5268}
  {arXiv:1404.5268} \BibitemShut {NoStop}%
\bibitem [{\citenamefont {WEN}(1990)}]{Wen1990}%
  \BibitemOpen
  \bibfield  {author} {\bibinfo {author} {\bibfnamefont {X.~G.}\ \bibnamefont
  {WEN}},\ }\href {\doibase 10.1142/s0217979290000139} {\bibfield  {journal}
  {\bibinfo  {journal} {International Journal of Modern Physics B}\ }\textbf
  {\bibinfo {volume} {04}},\ \bibinfo {pages} {239} (\bibinfo {year}
  {1990})}\BibitemShut {NoStop}%
\bibitem [{\citenamefont {Wen}(1995)}]{Wen1995}%
  \BibitemOpen
  \bibfield  {author} {\bibinfo {author} {\bibfnamefont {X.-G.}\ \bibnamefont
  {Wen}},\ }\href {\doibase 10.1080/00018739500101566} {\bibfield  {journal}
  {\bibinfo  {journal} {Advances in Physics}\ }\textbf {\bibinfo {volume}
  {44}},\ \bibinfo {pages} {405} (\bibinfo {year} {1995})},\ \Eprint
  {http://arxiv.org/abs/9506066} {arXiv:9506066 [cond-mat]} \BibitemShut
  {NoStop}%
\bibitem [{\citenamefont {Jahromi}\ \emph
  {et~al.}(2013{\natexlab{a}})\citenamefont {Jahromi}, \citenamefont {Masoudi},
  \citenamefont {Kargarian},\ and\ \citenamefont {Schmidt}}]{Jahromi2013a}%
  \BibitemOpen
  \bibfield  {author} {\bibinfo {author} {\bibfnamefont {S.~S.}\ \bibnamefont
  {Jahromi}}, \bibinfo {author} {\bibfnamefont {S.~F.}\ \bibnamefont
  {Masoudi}}, \bibinfo {author} {\bibfnamefont {M.}~\bibnamefont {Kargarian}},
  \ and\ \bibinfo {author} {\bibfnamefont {K.~P.}\ \bibnamefont {Schmidt}},\
  }\href {\doibase 10.1103/PhysRevB.88.214411} {\bibfield  {journal} {\bibinfo
  {journal} {Physical Review B - Condensed Matter and Materials Physics}\
  }\textbf {\bibinfo {volume} {88}},\ \bibinfo {pages} {214411} (\bibinfo
  {year} {2013}{\natexlab{a}})},\ \Eprint {http://arxiv.org/abs/1308.1407}
  {arXiv:1308.1407} \BibitemShut {NoStop}%
\bibitem [{\citenamefont {Jahromi}\ \emph
  {et~al.}(2013{\natexlab{b}})\citenamefont {Jahromi}, \citenamefont
  {Kargarian}, \citenamefont {Masoudi},\ and\ \citenamefont
  {Schmidt}}]{Jahromi2013}%
  \BibitemOpen
  \bibfield  {author} {\bibinfo {author} {\bibfnamefont {S.~S.}\ \bibnamefont
  {Jahromi}}, \bibinfo {author} {\bibfnamefont {M.}~\bibnamefont {Kargarian}},
  \bibinfo {author} {\bibfnamefont {S.~F.}\ \bibnamefont {Masoudi}}, \ and\
  \bibinfo {author} {\bibfnamefont {K.~P.}\ \bibnamefont {Schmidt}},\ }\href
  {\doibase 10.1103/PhysRevB.87.094413} {\bibfield  {journal} {\bibinfo
  {journal} {Physical Review B - Condensed Matter and Materials Physics}\
  }\textbf {\bibinfo {volume} {87}},\ \bibinfo {pages} {094413} (\bibinfo
  {year} {2013}{\natexlab{b}})},\ \Eprint {http://arxiv.org/abs/1211.1687}
  {arXiv:1211.1687} \BibitemShut {NoStop}%
\bibitem [{\citenamefont {Jahromi}\ \emph {et~al.}(2016)\citenamefont
  {Jahromi}, \citenamefont {Kargarian}, \citenamefont {Masoudi},\ and\
  \citenamefont {Langari}}]{Jahromi2016}%
  \BibitemOpen
  \bibfield  {author} {\bibinfo {author} {\bibfnamefont {S.~S.}\ \bibnamefont
  {Jahromi}}, \bibinfo {author} {\bibfnamefont {M.}~\bibnamefont {Kargarian}},
  \bibinfo {author} {\bibfnamefont {S.~F.}\ \bibnamefont {Masoudi}}, \ and\
  \bibinfo {author} {\bibfnamefont {A.}~\bibnamefont {Langari}},\ }\href
  {\doibase 10.1103/PhysRevB.94.125145} {\bibfield  {journal} {\bibinfo
  {journal} {Physical Review B}\ }\textbf {\bibinfo {volume} {94}},\ \bibinfo
  {pages} {125145} (\bibinfo {year} {2016})}\BibitemShut {NoStop}%
\bibitem [{\citenamefont {Mohseninia}\ \emph {et~al.}(2015)\citenamefont
  {Mohseninia}, \citenamefont {Jahromi}, \citenamefont {Memarzadeh},\ and\
  \citenamefont {Karimipour}}]{Mohseninia2015}%
  \BibitemOpen
  \bibfield  {author} {\bibinfo {author} {\bibfnamefont {R.}~\bibnamefont
  {Mohseninia}}, \bibinfo {author} {\bibfnamefont {S.~S.}\ \bibnamefont
  {Jahromi}}, \bibinfo {author} {\bibfnamefont {L.}~\bibnamefont {Memarzadeh}},
  \ and\ \bibinfo {author} {\bibfnamefont {V.}~\bibnamefont {Karimipour}},\
  }\href {\doibase 10.1103/PhysRevB.91.245110} {\bibfield  {journal} {\bibinfo
  {journal} {Physical Review B - Condensed Matter and Materials Physics}\
  }\textbf {\bibinfo {volume} {91}},\ \bibinfo {pages} {245110} (\bibinfo
  {year} {2015})},\ \Eprint {http://arxiv.org/abs/1503.05957}
  {arXiv:1503.05957} \BibitemShut {NoStop}%
\bibitem [{\citenamefont {Capponi}\ \emph {et~al.}(2014)\citenamefont
  {Capponi}, \citenamefont {Jahromi}, \citenamefont {Alet},\ and\ \citenamefont
  {Schmidt}}]{Capponi2014}%
  \BibitemOpen
  \bibfield  {author} {\bibinfo {author} {\bibfnamefont {S.}~\bibnamefont
  {Capponi}}, \bibinfo {author} {\bibfnamefont {S.~S.}\ \bibnamefont
  {Jahromi}}, \bibinfo {author} {\bibfnamefont {F.}~\bibnamefont {Alet}}, \
  and\ \bibinfo {author} {\bibfnamefont {K.~P.}\ \bibnamefont {Schmidt}},\
  }\href {\doibase 10.1103/PhysRevE.89.062136} {\bibfield  {journal} {\bibinfo
  {journal} {Physical Review E - Statistical, Nonlinear, and Soft Matter
  Physics}\ }\textbf {\bibinfo {volume} {89}},\ \bibinfo {pages} {062136}
  (\bibinfo {year} {2014})},\ \Eprint {http://arxiv.org/abs/1403.1406}
  {arXiv:1403.1406} \BibitemShut {NoStop}%
\bibitem [{\citenamefont {Khaliullin}(2005)}]{Khaliullin2005}%
  \BibitemOpen
  \bibfield  {author} {\bibinfo {author} {\bibfnamefont {G.}~\bibnamefont
  {Khaliullin}},\ }\href {\doibase 10.1143/PTPS.160.155} {\bibfield  {journal}
  {\bibinfo  {journal} {Progress of Theoretical Physics Supplement}\ }\textbf
  {\bibinfo {volume} {160}},\ \bibinfo {pages} {155} (\bibinfo {year}
  {2005})}\BibitemShut {NoStop}%
\bibitem [{\citenamefont {Trebst}(2017)}]{Trebst2017}%
  \BibitemOpen
  \bibfield  {author} {\bibinfo {author} {\bibfnamefont {S.}~\bibnamefont
  {Trebst}},\ }\href {http://arxiv.org/abs/1701.07056} {\  (\bibinfo {year}
  {2017})},\ \Eprint {http://arxiv.org/abs/1701.07056} {arXiv:1701.07056}
  \BibitemShut {NoStop}%
\bibitem [{\citenamefont {Pesin}\ and\ \citenamefont
  {Balents}(2010)}]{Pesin2010}%
  \BibitemOpen
  \bibfield  {author} {\bibinfo {author} {\bibfnamefont {D.}~\bibnamefont
  {Pesin}}\ and\ \bibinfo {author} {\bibfnamefont {L.}~\bibnamefont
  {Balents}},\ }\href {\doibase 10.1038/nphys1606} {\bibfield  {journal}
  {\bibinfo  {journal} {Nature Physics}\ }\textbf {\bibinfo {volume} {6}},\
  \bibinfo {pages} {376} (\bibinfo {year} {2010})}\BibitemShut {NoStop}%
\bibitem [{\citenamefont {Kasahara}\ \emph {et~al.}(2018)\citenamefont
  {Kasahara}, \citenamefont {Ohnishi}, \citenamefont {Mizukami}, \citenamefont
  {Tanaka}, \citenamefont {Ma}, \citenamefont {Sugii}, \citenamefont {Kurita},
  \citenamefont {Tanaka}, \citenamefont {Nasu}, \citenamefont {Motome},
  \citenamefont {Shibauchi},\ and\ \citenamefont {Matsuda}}]{Kasahara2018}%
  \BibitemOpen
  \bibfield  {author} {\bibinfo {author} {\bibfnamefont {Y.}~\bibnamefont
  {Kasahara}}, \bibinfo {author} {\bibfnamefont {T.}~\bibnamefont {Ohnishi}},
  \bibinfo {author} {\bibfnamefont {Y.}~\bibnamefont {Mizukami}}, \bibinfo
  {author} {\bibfnamefont {O.}~\bibnamefont {Tanaka}}, \bibinfo {author}
  {\bibfnamefont {S.}~\bibnamefont {Ma}}, \bibinfo {author} {\bibfnamefont
  {K.}~\bibnamefont {Sugii}}, \bibinfo {author} {\bibfnamefont
  {N.}~\bibnamefont {Kurita}}, \bibinfo {author} {\bibfnamefont
  {H.}~\bibnamefont {Tanaka}}, \bibinfo {author} {\bibfnamefont
  {J.}~\bibnamefont {Nasu}}, \bibinfo {author} {\bibfnamefont {Y.}~\bibnamefont
  {Motome}}, \bibinfo {author} {\bibfnamefont {T.}~\bibnamefont {Shibauchi}}, \
  and\ \bibinfo {author} {\bibfnamefont {Y.}~\bibnamefont {Matsuda}},\ }\href
  {\doibase 10.1038/s41586-018-0274-0} {\bibfield  {journal} {\bibinfo
  {journal} {Nature}\ }\textbf {\bibinfo {volume} {559}},\ \bibinfo {pages}
  {227} (\bibinfo {year} {2018})}\BibitemShut {NoStop}%
\bibitem [{\citenamefont {O'Brien}\ \emph {et~al.}(2016)\citenamefont
  {O'Brien}, \citenamefont {Hermanns},\ and\ \citenamefont
  {Trebst}}]{OBrien2016}%
  \BibitemOpen
  \bibfield  {author} {\bibinfo {author} {\bibfnamefont {K.}~\bibnamefont
  {O'Brien}}, \bibinfo {author} {\bibfnamefont {M.}~\bibnamefont {Hermanns}}, \
  and\ \bibinfo {author} {\bibfnamefont {S.}~\bibnamefont {Trebst}},\ }\href
  {\doibase 10.1103/PhysRevB.93.085101} {\bibfield  {journal} {\bibinfo
  {journal} {Physical Review B}\ }\textbf {\bibinfo {volume} {93}},\ \bibinfo
  {pages} {085101} (\bibinfo {year} {2016})},\ \Eprint
  {http://arxiv.org/abs/1511.05569} {arXiv:1511.05569} \BibitemShut {NoStop}%
\bibitem [{\citenamefont {Mandal}\ and\ \citenamefont
  {Surendran}(2009)}]{Mandal2009}%
  \BibitemOpen
  \bibfield  {author} {\bibinfo {author} {\bibfnamefont {S.}~\bibnamefont
  {Mandal}}\ and\ \bibinfo {author} {\bibfnamefont {N.}~\bibnamefont
  {Surendran}},\ }\href {\doibase 10.1103/PhysRevB.79.024426} {\bibfield
  {journal} {\bibinfo  {journal} {Physical Review B - Condensed Matter and
  Materials Physics}\ }\textbf {\bibinfo {volume} {79}},\ \bibinfo {pages}
  {024426} (\bibinfo {year} {2009})},\ \Eprint {http://arxiv.org/abs/0801.0229}
  {arXiv:0801.0229} \BibitemShut {NoStop}%
\bibitem [{\citenamefont {Si}\ and\ \citenamefont {Yu}(2008)}]{SI2008428}%
  \BibitemOpen
  \bibfield  {author} {\bibinfo {author} {\bibfnamefont {T.}~\bibnamefont
  {Si}}\ and\ \bibinfo {author} {\bibfnamefont {Y.}~\bibnamefont {Yu}},\ }\href
  {\doibase https://doi.org/10.1016/j.nuclphysb.2008.06.009} {\bibfield
  {journal} {\bibinfo  {journal} {Nuclear Physics B}\ }\textbf {\bibinfo
  {volume} {803}},\ \bibinfo {pages} {428} (\bibinfo {year}
  {2008})}\BibitemShut {NoStop}%
\bibitem [{\citenamefont {Hermanns}\ \emph {et~al.}(2015)\citenamefont
  {Hermanns}, \citenamefont {O'Brien},\ and\ \citenamefont
  {Trebst}}]{Hermanns2015}%
  \BibitemOpen
  \bibfield  {author} {\bibinfo {author} {\bibfnamefont {M.}~\bibnamefont
  {Hermanns}}, \bibinfo {author} {\bibfnamefont {K.}~\bibnamefont {O'Brien}}, \
  and\ \bibinfo {author} {\bibfnamefont {S.}~\bibnamefont {Trebst}},\ }\href
  {\doibase 10.1103/PhysRevLett.114.157202} {\bibfield  {journal} {\bibinfo
  {journal} {Physical Review Letters}\ }\textbf {\bibinfo {volume} {114}},\
  \bibinfo {pages} {157202} (\bibinfo {year} {2015})},\ \Eprint
  {http://arxiv.org/abs/1411.7379} {arXiv:1411.7379} \BibitemShut {NoStop}%
\bibitem [{\citenamefont {Takayama}\ \emph {et~al.}(2015)\citenamefont
  {Takayama}, \citenamefont {Kato}, \citenamefont {Dinnebier}, \citenamefont
  {Nuss}, \citenamefont {Kono}, \citenamefont {Veiga}, \citenamefont {Fabbris},
  \citenamefont {Haskel},\ and\ \citenamefont {Takagi}}]{Takayama2015}%
  \BibitemOpen
  \bibfield  {author} {\bibinfo {author} {\bibfnamefont {T.}~\bibnamefont
  {Takayama}}, \bibinfo {author} {\bibfnamefont {A.}~\bibnamefont {Kato}},
  \bibinfo {author} {\bibfnamefont {R.}~\bibnamefont {Dinnebier}}, \bibinfo
  {author} {\bibfnamefont {J.}~\bibnamefont {Nuss}}, \bibinfo {author}
  {\bibfnamefont {H.}~\bibnamefont {Kono}}, \bibinfo {author} {\bibfnamefont
  {L.~S.}\ \bibnamefont {Veiga}}, \bibinfo {author} {\bibfnamefont
  {G.}~\bibnamefont {Fabbris}}, \bibinfo {author} {\bibfnamefont
  {D.}~\bibnamefont {Haskel}}, \ and\ \bibinfo {author} {\bibfnamefont
  {H.}~\bibnamefont {Takagi}},\ }\href {\doibase
  10.1103/PhysRevLett.114.077202} {\bibfield  {journal} {\bibinfo  {journal}
  {Physical Review Letters}\ }\textbf {\bibinfo {volume} {114}},\ \bibinfo
  {pages} {077202} (\bibinfo {year} {2015})}\BibitemShut {NoStop}%
\bibitem [{\citenamefont {Hermanns}\ and\ \citenamefont
  {Trebst}(2014)}]{Hermanns2014}%
  \BibitemOpen
  \bibfield  {author} {\bibinfo {author} {\bibfnamefont {M.}~\bibnamefont
  {Hermanns}}\ and\ \bibinfo {author} {\bibfnamefont {S.}~\bibnamefont
  {Trebst}},\ }\href {\doibase 10.1103/PhysRevB.89.235102} {\bibfield
  {journal} {\bibinfo  {journal} {Physical Review B - Condensed Matter and
  Materials Physics}\ }\textbf {\bibinfo {volume} {89}},\ \bibinfo {pages}
  {235102} (\bibinfo {year} {2014})},\ \Eprint {http://arxiv.org/abs/1401.7678}
  {arXiv:1401.7678} \BibitemShut {NoStop}%
\bibitem [{\citenamefont {Kargarian}\ and\ \citenamefont
  {Fiete}(2010)}]{Kargarian2010}%
  \BibitemOpen
  \bibfield  {author} {\bibinfo {author} {\bibfnamefont {M.}~\bibnamefont
  {Kargarian}}\ and\ \bibinfo {author} {\bibfnamefont {G.~A.}\ \bibnamefont
  {Fiete}},\ }\href {\doibase 10.1103/PhysRevB.82.085106} {\bibfield  {journal}
  {\bibinfo  {journal} {Physical Review B - Condensed Matter and Materials
  Physics}\ }\textbf {\bibinfo {volume} {82}},\ \bibinfo {pages} {085106}
  (\bibinfo {year} {2010})},\ \Eprint {http://arxiv.org/abs/1005.3815}
  {arXiv:1005.3815} \BibitemShut {NoStop}%
\bibitem [{\citenamefont {Modic}\ \emph {et~al.}(2014)\citenamefont {Modic},
  \citenamefont {Smidt}, \citenamefont {Kimchi}, \citenamefont {Breznay},
  \citenamefont {Biffin}, \citenamefont {Choi}, \citenamefont {Johnson},
  \citenamefont {Coldea}, \citenamefont {Watkins-Curry}, \citenamefont
  {McCandless}, \citenamefont {Chan}, \citenamefont {Gandara}, \citenamefont
  {Islam}, \citenamefont {Vishwanath}, \citenamefont {Shekhter}, \citenamefont
  {McDonald},\ and\ \citenamefont {Analytis}}]{Modic2014}%
  \BibitemOpen
  \bibfield  {author} {\bibinfo {author} {\bibfnamefont {K.~A.}\ \bibnamefont
  {Modic}}, \bibinfo {author} {\bibfnamefont {T.~E.}\ \bibnamefont {Smidt}},
  \bibinfo {author} {\bibfnamefont {I.}~\bibnamefont {Kimchi}}, \bibinfo
  {author} {\bibfnamefont {N.~P.}\ \bibnamefont {Breznay}}, \bibinfo {author}
  {\bibfnamefont {A.}~\bibnamefont {Biffin}}, \bibinfo {author} {\bibfnamefont
  {S.}~\bibnamefont {Choi}}, \bibinfo {author} {\bibfnamefont {R.~D.}\
  \bibnamefont {Johnson}}, \bibinfo {author} {\bibfnamefont {R.}~\bibnamefont
  {Coldea}}, \bibinfo {author} {\bibfnamefont {P.}~\bibnamefont
  {Watkins-Curry}}, \bibinfo {author} {\bibfnamefont {G.~T.}\ \bibnamefont
  {McCandless}}, \bibinfo {author} {\bibfnamefont {J.~Y.}\ \bibnamefont
  {Chan}}, \bibinfo {author} {\bibfnamefont {F.}~\bibnamefont {Gandara}},
  \bibinfo {author} {\bibfnamefont {Z.}~\bibnamefont {Islam}}, \bibinfo
  {author} {\bibfnamefont {A.}~\bibnamefont {Vishwanath}}, \bibinfo {author}
  {\bibfnamefont {A.}~\bibnamefont {Shekhter}}, \bibinfo {author}
  {\bibfnamefont {R.~D.}\ \bibnamefont {McDonald}}, \ and\ \bibinfo {author}
  {\bibfnamefont {J.~G.}\ \bibnamefont {Analytis}},\ }\href {\doibase
  10.1038/ncomms5203} {\bibfield  {journal} {\bibinfo  {journal} {Nature
  Communications}\ }\textbf {\bibinfo {volume} {5}},\ \bibinfo {pages} {1}
  (\bibinfo {year} {2014})}\BibitemShut {NoStop}%
\bibitem [{\citenamefont {Chaloupka}\ \emph {et~al.}(2010)\citenamefont
  {Chaloupka}, \citenamefont {Jackeli},\ and\ \citenamefont
  {Khaliullin}}]{Chaloupka2010}%
  \BibitemOpen
  \bibfield  {author} {\bibinfo {author} {\bibfnamefont {J.}~\bibnamefont
  {Chaloupka}}, \bibinfo {author} {\bibfnamefont {G.}~\bibnamefont {Jackeli}},
  \ and\ \bibinfo {author} {\bibfnamefont {G.}~\bibnamefont {Khaliullin}},\
  }\href {\doibase 10.1103/PhysRevLett.105.027204} {\bibfield  {journal}
  {\bibinfo  {journal} {Physical Review Letters}\ }\textbf {\bibinfo {volume}
  {105}},\ \bibinfo {pages} {027204} (\bibinfo {year} {2010})}\BibitemShut
  {NoStop}%
\bibitem [{\citenamefont {Chaloupka}\ \emph {et~al.}(2013)\citenamefont
  {Chaloupka}, \citenamefont {Jackeli},\ and\ \citenamefont
  {Khaliullin}}]{Chaloupka2013}%
  \BibitemOpen
  \bibfield  {author} {\bibinfo {author} {\bibfnamefont {J.}~\bibnamefont
  {Chaloupka}}, \bibinfo {author} {\bibfnamefont {G.}~\bibnamefont {Jackeli}},
  \ and\ \bibinfo {author} {\bibfnamefont {G.}~\bibnamefont {Khaliullin}},\
  }\href {\doibase 10.1103/PhysRevLett.110.097204} {\bibfield  {journal}
  {\bibinfo  {journal} {Physical Review Letters}\ }\textbf {\bibinfo {volume}
  {110}},\ \bibinfo {pages} {097204} (\bibinfo {year} {2013})}\BibitemShut
  {NoStop}%
\bibitem [{\citenamefont {Morita}\ and\ \citenamefont
  {Tohyama}(2020)}]{Morita2020}%
  \BibitemOpen
  \bibfield  {author} {\bibinfo {author} {\bibfnamefont {K.}~\bibnamefont
  {Morita}}\ and\ \bibinfo {author} {\bibfnamefont {T.}~\bibnamefont
  {Tohyama}},\ }\href {\doibase 10.1103/physrevresearch.2.013205} {\bibfield
  {journal} {\bibinfo  {journal} {Physical Review Research}\ }\textbf {\bibinfo
  {volume} {2}},\ \bibinfo {pages} {013205} (\bibinfo {year} {2020})},\ \Eprint
  {http://arxiv.org/abs/1911.09266} {arXiv:1911.09266} \BibitemShut {NoStop}%
\bibitem [{\citenamefont {Hickey}\ and\ \citenamefont
  {Trebst}(2019)}]{Hickey2019}%
  \BibitemOpen
  \bibfield  {author} {\bibinfo {author} {\bibfnamefont {C.}~\bibnamefont
  {Hickey}}\ and\ \bibinfo {author} {\bibfnamefont {S.}~\bibnamefont
  {Trebst}},\ }\href {\doibase 10.1038/s41467-019-08459-9} {\bibfield
  {journal} {\bibinfo  {journal} {Nature Communications}\ }\textbf {\bibinfo
  {volume} {10}},\ \bibinfo {pages} {530} (\bibinfo {year} {2019})}\BibitemShut
  {NoStop}%
\bibitem [{\citenamefont {Koga}\ \emph {et~al.}(2018)\citenamefont {Koga},
  \citenamefont {Tomishige},\ and\ \citenamefont {Nasu}}]{Koga2018}%
  \BibitemOpen
  \bibfield  {author} {\bibinfo {author} {\bibfnamefont {A.}~\bibnamefont
  {Koga}}, \bibinfo {author} {\bibfnamefont {H.}~\bibnamefont {Tomishige}}, \
  and\ \bibinfo {author} {\bibfnamefont {J.}~\bibnamefont {Nasu}},\ }\href
  {\doibase 10.7566/JPSJ.87.063703} {\bibfield  {journal} {\bibinfo  {journal}
  {Journal of the Physical Society of Japan}\ }\textbf {\bibinfo {volume} {87}}
  (\bibinfo {year} {2018}),\ 10.7566/JPSJ.87.063703},\ \Eprint
  {http://arxiv.org/abs/1803.08385} {arXiv:1803.08385} \BibitemShut {NoStop}%
\bibitem [{\citenamefont {Suzuki}\ and\ \citenamefont
  {Yamaji}(2018)}]{SUZUKI2018637}%
  \BibitemOpen
  \bibfield  {author} {\bibinfo {author} {\bibfnamefont {T.}~\bibnamefont
  {Suzuki}}\ and\ \bibinfo {author} {\bibfnamefont {Y.}~\bibnamefont
  {Yamaji}},\ }\href {\doibase https://doi.org/10.1016/j.physb.2017.09.105}
  {\bibfield  {journal} {\bibinfo  {journal} {Physica B: Condensed Matter}\
  }\textbf {\bibinfo {volume} {536}},\ \bibinfo {pages} {637} (\bibinfo {year}
  {2018})}\BibitemShut {NoStop}%
\bibitem [{\citenamefont {Nasu}\ \emph {et~al.}(2015)\citenamefont {Nasu},
  \citenamefont {Udagawa},\ and\ \citenamefont {Motome}}]{Nasu2015}%
  \BibitemOpen
  \bibfield  {author} {\bibinfo {author} {\bibfnamefont {J.}~\bibnamefont
  {Nasu}}, \bibinfo {author} {\bibfnamefont {M.}~\bibnamefont {Udagawa}}, \
  and\ \bibinfo {author} {\bibfnamefont {Y.}~\bibnamefont {Motome}},\ }\href
  {\doibase 10.1103/PhysRevB.92.115122} {\bibfield  {journal} {\bibinfo
  {journal} {Physical Review B - Condensed Matter and Materials Physics}\
  }\textbf {\bibinfo {volume} {92}},\ \bibinfo {pages} {115122} (\bibinfo
  {year} {2015})},\ \Eprint {http://arxiv.org/abs/1504.01259}
  {arXiv:1504.01259} \BibitemShut {NoStop}%
\bibitem [{\citenamefont {Czarnik}\ \emph
  {et~al.}(2019{\natexlab{a}})\citenamefont {Czarnik}, \citenamefont
  {Francuz},\ and\ \citenamefont {Dziarmaga}}]{Czarnik2019}%
  \BibitemOpen
  \bibfield  {author} {\bibinfo {author} {\bibfnamefont {P.}~\bibnamefont
  {Czarnik}}, \bibinfo {author} {\bibfnamefont {A.}~\bibnamefont {Francuz}}, \
  and\ \bibinfo {author} {\bibfnamefont {J.}~\bibnamefont {Dziarmaga}},\ }\href
  {\doibase 10.1103/PhysRevB.100.165147} {\bibfield  {journal} {\bibinfo
  {journal} {Physical Review B}\ }\textbf {\bibinfo {volume} {100}},\ \bibinfo
  {pages} {165147} (\bibinfo {year} {2019}{\natexlab{a}})},\ \Eprint
  {http://arxiv.org/abs/1906.02220} {arXiv:1906.02220} \BibitemShut {NoStop}%
\bibitem [{\citenamefont {Lee}\ \emph {et~al.}(2014)\citenamefont {Lee},
  \citenamefont {Schaffer}, \citenamefont {Bhattacharjee},\ and\ \citenamefont
  {Kim}}]{Lee2014}%
  \BibitemOpen
  \bibfield  {author} {\bibinfo {author} {\bibfnamefont {E.~K.~H.}\
  \bibnamefont {Lee}}, \bibinfo {author} {\bibfnamefont {R.}~\bibnamefont
  {Schaffer}}, \bibinfo {author} {\bibfnamefont {S.}~\bibnamefont
  {Bhattacharjee}}, \ and\ \bibinfo {author} {\bibfnamefont {Y.~B.}\
  \bibnamefont {Kim}},\ }\href {\doibase 10.1103/PhysRevB.89.045117} {\bibfield
   {journal} {\bibinfo  {journal} {Physical Review B - Condensed Matter and
  Materials Physics}\ }\textbf {\bibinfo {volume} {89}},\ \bibinfo {pages}
  {045117} (\bibinfo {year} {2014})},\ \Eprint {http://arxiv.org/abs/1308.6592}
  {arXiv:1308.6592} \BibitemShut {NoStop}%
\bibitem [{\citenamefont {Singh}\ and\ \citenamefont
  {Oitmaa}(2017)}]{Singh2017}%
  \BibitemOpen
  \bibfield  {author} {\bibinfo {author} {\bibfnamefont {R.~R.}\ \bibnamefont
  {Singh}}\ and\ \bibinfo {author} {\bibfnamefont {J.}~\bibnamefont {Oitmaa}},\
  }\href {\doibase 10.1103/PhysRevB.96.144414} {\bibfield  {journal} {\bibinfo
  {journal} {Physical Review B}\ }\textbf {\bibinfo {volume} {96}},\ \bibinfo
  {pages} {144414} (\bibinfo {year} {2017})},\ \Eprint
  {http://arxiv.org/abs/1707.01126} {arXiv:1707.01126} \BibitemShut {NoStop}%
\bibitem [{\citenamefont {Eschmann}\ \emph {et~al.}(2020)\citenamefont
  {Eschmann}, \citenamefont {Mishchenko}, \citenamefont {O'Brien},
  \citenamefont {Bojesen}, \citenamefont {Kato}, \citenamefont {Hermanns},
  \citenamefont {Motome},\ and\ \citenamefont {Trebst}}]{Eschmann2020}%
  \BibitemOpen
  \bibfield  {author} {\bibinfo {author} {\bibfnamefont {T.}~\bibnamefont
  {Eschmann}}, \bibinfo {author} {\bibfnamefont {P.~A.}\ \bibnamefont
  {Mishchenko}}, \bibinfo {author} {\bibfnamefont {K.}~\bibnamefont {O'Brien}},
  \bibinfo {author} {\bibfnamefont {T.~A.}\ \bibnamefont {Bojesen}}, \bibinfo
  {author} {\bibfnamefont {Y.}~\bibnamefont {Kato}}, \bibinfo {author}
  {\bibfnamefont {M.}~\bibnamefont {Hermanns}}, \bibinfo {author}
  {\bibfnamefont {Y.}~\bibnamefont {Motome}}, \ and\ \bibinfo {author}
  {\bibfnamefont {S.}~\bibnamefont {Trebst}},\ }\href {\doibase
  10.1103/PhysRevB.102.075125} {\bibfield  {journal} {\bibinfo  {journal}
  {Physical Review B}\ }\textbf {\bibinfo {volume} {102}},\ \bibinfo {pages}
  {075125} (\bibinfo {year} {2020})},\ \Eprint
  {http://arxiv.org/abs/2006.07386} {arXiv:2006.07386} \BibitemShut {NoStop}%
\bibitem [{\citenamefont {Mishchenko}\ \emph {et~al.}(2017)\citenamefont
  {Mishchenko}, \citenamefont {Kato},\ and\ \citenamefont
  {Motome}}]{Mishchenko2017}%
  \BibitemOpen
  \bibfield  {author} {\bibinfo {author} {\bibfnamefont {P.~A.}\ \bibnamefont
  {Mishchenko}}, \bibinfo {author} {\bibfnamefont {Y.}~\bibnamefont {Kato}}, \
  and\ \bibinfo {author} {\bibfnamefont {Y.}~\bibnamefont {Motome}},\ }\href
  {\doibase 10.1103/PhysRevB.96.125124} {\bibfield  {journal} {\bibinfo
  {journal} {Physical Review B}\ }\textbf {\bibinfo {volume} {96}},\ \bibinfo
  {pages} {125124} (\bibinfo {year} {2017})},\ \Eprint
  {http://arxiv.org/abs/1706.05057} {arXiv:1706.05057} \BibitemShut {NoStop}%
\bibitem [{\citenamefont {Nasu}\ \emph
  {et~al.}(2014{\natexlab{a}})\citenamefont {Nasu}, \citenamefont {Kaji},
  \citenamefont {Matsuura}, \citenamefont {Udagawa},\ and\ \citenamefont
  {Motome}}]{Nasu2014}%
  \BibitemOpen
  \bibfield  {author} {\bibinfo {author} {\bibfnamefont {J.}~\bibnamefont
  {Nasu}}, \bibinfo {author} {\bibfnamefont {T.}~\bibnamefont {Kaji}}, \bibinfo
  {author} {\bibfnamefont {K.}~\bibnamefont {Matsuura}}, \bibinfo {author}
  {\bibfnamefont {M.}~\bibnamefont {Udagawa}}, \ and\ \bibinfo {author}
  {\bibfnamefont {Y.}~\bibnamefont {Motome}},\ }\href {\doibase
  10.1103/PhysRevB.89.115125} {\bibfield  {journal} {\bibinfo  {journal}
  {Physical Review B - Condensed Matter and Materials Physics}\ }\textbf
  {\bibinfo {volume} {89}},\ \bibinfo {pages} {115125} (\bibinfo {year}
  {2014}{\natexlab{a}})},\ \Eprint {http://arxiv.org/abs/1309.3068}
  {arXiv:1309.3068} \BibitemShut {NoStop}%
\bibitem [{\citenamefont {Nasu}\ \emph
  {et~al.}(2014{\natexlab{b}})\citenamefont {Nasu}, \citenamefont {Udagawa},\
  and\ \citenamefont {Motome}}]{Nasu2014a}%
  \BibitemOpen
  \bibfield  {author} {\bibinfo {author} {\bibfnamefont {J.}~\bibnamefont
  {Nasu}}, \bibinfo {author} {\bibfnamefont {M.}~\bibnamefont {Udagawa}}, \
  and\ \bibinfo {author} {\bibfnamefont {Y.}~\bibnamefont {Motome}},\ }\href
  {\doibase 10.1103/PhysRevLett.113.197205} {\bibfield  {journal} {\bibinfo
  {journal} {Physical Review Letters}\ }\textbf {\bibinfo {volume} {113}},\
  \bibinfo {pages} {197205} (\bibinfo {year} {2014}{\natexlab{b}})},\ \Eprint
  {http://arxiv.org/abs/1406.5415} {arXiv:1406.5415} \BibitemShut {NoStop}%
\bibitem [{\citenamefont {Corboz}\ \emph {et~al.}(2014)\citenamefont {Corboz},
  \citenamefont {Rice},\ and\ \citenamefont {Troyer}}]{Corboz2014a}%
  \BibitemOpen
  \bibfield  {author} {\bibinfo {author} {\bibfnamefont {P.}~\bibnamefont
  {Corboz}}, \bibinfo {author} {\bibfnamefont {T.~M.}\ \bibnamefont {Rice}}, \
  and\ \bibinfo {author} {\bibfnamefont {M.}~\bibnamefont {Troyer}},\ }\href
  {\doibase 10.1103/PhysRevLett.113.046402} {\bibfield  {journal} {\bibinfo
  {journal} {Physical Review Letters}\ }\textbf {\bibinfo {volume} {113}},\
  \bibinfo {pages} {046402} (\bibinfo {year} {2014})},\ \Eprint
  {http://arxiv.org/abs/1402.2859} {arXiv:1402.2859} \BibitemShut {NoStop}%
\bibitem [{\citenamefont {Corboz}\ \emph {et~al.}(2012)\citenamefont {Corboz},
  \citenamefont {Penc}, \citenamefont {Mila},\ and\ \citenamefont
  {L{\"{a}}uchli}}]{Corboz2012a}%
  \BibitemOpen
  \bibfield  {author} {\bibinfo {author} {\bibfnamefont {P.}~\bibnamefont
  {Corboz}}, \bibinfo {author} {\bibfnamefont {K.}~\bibnamefont {Penc}},
  \bibinfo {author} {\bibfnamefont {F.}~\bibnamefont {Mila}}, \ and\ \bibinfo
  {author} {\bibfnamefont {A.~M.}\ \bibnamefont {L{\"{a}}uchli}},\ }\href
  {\doibase 10.1103/PhysRevB.86.041106} {\bibfield  {journal} {\bibinfo
  {journal} {Physical Review B - Condensed Matter and Materials Physics}\
  }\textbf {\bibinfo {volume} {86}},\ \bibinfo {pages} {041106} (\bibinfo
  {year} {2012})},\ \Eprint {http://arxiv.org/abs/1204.6682} {arXiv:1204.6682}
  \BibitemShut {NoStop}%
\bibitem [{\citenamefont {Corboz}\ and\ \citenamefont
  {Mila}(2013)}]{Corboz2013}%
  \BibitemOpen
  \bibfield  {author} {\bibinfo {author} {\bibfnamefont {P.}~\bibnamefont
  {Corboz}}\ and\ \bibinfo {author} {\bibfnamefont {F.}~\bibnamefont {Mila}},\
  }\href {\doibase 10.1103/PhysRevB.87.115144} {\bibfield  {journal} {\bibinfo
  {journal} {Physical Review B - Condensed Matter and Materials Physics}\
  }\textbf {\bibinfo {volume} {87}},\ \bibinfo {pages} {115144} (\bibinfo
  {year} {2013})},\ \Eprint {http://arxiv.org/abs/1212.2983} {arXiv:1212.2983}
  \BibitemShut {NoStop}%
\bibitem [{\citenamefont {Corboz}\ and\ \citenamefont
  {Mila}(2014)}]{Corboz2014}%
  \BibitemOpen
  \bibfield  {author} {\bibinfo {author} {\bibfnamefont {P.}~\bibnamefont
  {Corboz}}\ and\ \bibinfo {author} {\bibfnamefont {F.}~\bibnamefont {Mila}},\
  }\href {\doibase 10.1103/PhysRevLett.112.147203} {\bibfield  {journal}
  {\bibinfo  {journal} {Physical Review Letters}\ }\textbf {\bibinfo {volume}
  {112}},\ \bibinfo {pages} {147203} (\bibinfo {year} {2014})},\ \Eprint
  {http://arxiv.org/abs/arXiv:1401.3778v1} {arXiv:arXiv:1401.3778v1}
  \BibitemShut {NoStop}%
\bibitem [{\citenamefont {Jahromi}\ and\ \citenamefont
  {Or{\'{u}}s}(2018)}]{Jahromi2018}%
  \BibitemOpen
  \bibfield  {author} {\bibinfo {author} {\bibfnamefont {S.~S.}\ \bibnamefont
  {Jahromi}}\ and\ \bibinfo {author} {\bibfnamefont {R.}~\bibnamefont
  {Or{\'{u}}s}},\ }\href {\doibase 10.1103/PhysRevB.98.155108} {\bibfield
  {journal} {\bibinfo  {journal} {Physical Review B}\ }\textbf {\bibinfo
  {volume} {98}},\ \bibinfo {pages} {155108} (\bibinfo {year}
  {2018})}\BibitemShut {NoStop}%
\bibitem [{\citenamefont {Schmoll}\ \emph {et~al.}(2020)\citenamefont
  {Schmoll}, \citenamefont {Jahromi}, \citenamefont {H{\"{o}}rmann},
  \citenamefont {M{\"{u}}hlhauser}, \citenamefont {Schmidt},\ and\
  \citenamefont {Or{\'{u}}s}}]{Schmoll2020}%
  \BibitemOpen
  \bibfield  {author} {\bibinfo {author} {\bibfnamefont {P.}~\bibnamefont
  {Schmoll}}, \bibinfo {author} {\bibfnamefont {S.~S.}\ \bibnamefont
  {Jahromi}}, \bibinfo {author} {\bibfnamefont {M.}~\bibnamefont
  {H{\"{o}}rmann}}, \bibinfo {author} {\bibfnamefont {M.}~\bibnamefont
  {M{\"{u}}hlhauser}}, \bibinfo {author} {\bibfnamefont {K.~P.}\ \bibnamefont
  {Schmidt}}, \ and\ \bibinfo {author} {\bibfnamefont {R.}~\bibnamefont
  {Or{\'{u}}s}},\ }\href {\doibase 10.1103/PhysRevLett.124.200603} {\bibfield
  {journal} {\bibinfo  {journal} {Physical Review Letters}\ }\textbf {\bibinfo
  {volume} {124}},\ \bibinfo {pages} {200603} (\bibinfo {year} {2020})},\
  \Eprint {http://arxiv.org/abs/1911.04882} {arXiv:1911.04882} \BibitemShut
  {NoStop}%
\bibitem [{\citenamefont {Sadrzadeh}\ \emph {et~al.}(2016)\citenamefont
  {Sadrzadeh}, \citenamefont {Haghshenas}, \citenamefont {Jahromi},\ and\
  \citenamefont {Langari}}]{Sadrzadeh2016}%
  \BibitemOpen
  \bibfield  {author} {\bibinfo {author} {\bibfnamefont {M.}~\bibnamefont
  {Sadrzadeh}}, \bibinfo {author} {\bibfnamefont {R.}~\bibnamefont
  {Haghshenas}}, \bibinfo {author} {\bibfnamefont {S.~S.}\ \bibnamefont
  {Jahromi}}, \ and\ \bibinfo {author} {\bibfnamefont {A.}~\bibnamefont
  {Langari}},\ }\href {\doibase 10.1103/PhysRevB.94.214419} {\bibfield
  {journal} {\bibinfo  {journal} {Physical Review B}\ }\textbf {\bibinfo
  {volume} {94}},\ \bibinfo {pages} {214419} (\bibinfo {year} {2016})},\
  \Eprint {http://arxiv.org/abs/1611.08298} {arXiv:1611.08298} \BibitemShut
  {NoStop}%
\bibitem [{\citenamefont {Jahromi}\ and\ \citenamefont
  {Or{\'{u}}s}(2020)}]{Jahromi2020}%
  \BibitemOpen
  \bibfield  {author} {\bibinfo {author} {\bibfnamefont {S.~S.}\ \bibnamefont
  {Jahromi}}\ and\ \bibinfo {author} {\bibfnamefont {R.}~\bibnamefont
  {Or{\'{u}}s}},\ }\href {\doibase 10.1103/PhysRevB.101.115114} {\bibfield
  {journal} {\bibinfo  {journal} {Physical Review B}\ }\textbf {\bibinfo
  {volume} {101}},\ \bibinfo {pages} {115114} (\bibinfo {year}
  {2020})}\BibitemShut {NoStop}%
\bibitem [{\citenamefont {Jahromi}\ and\ \citenamefont
  {Or{\'{u}}s}(2019)}]{Jahromi2019}%
  \BibitemOpen
  \bibfield  {author} {\bibinfo {author} {\bibfnamefont {S.~S.}\ \bibnamefont
  {Jahromi}}\ and\ \bibinfo {author} {\bibfnamefont {R.}~\bibnamefont
  {Or{\'{u}}s}},\ }\href {\doibase 10.1103/PhysRevB.99.195105} {\bibfield
  {journal} {\bibinfo  {journal} {Physical Review B}\ }\textbf {\bibinfo
  {volume} {99}},\ \bibinfo {pages} {195105} (\bibinfo {year} {2019})},\
  \Eprint {http://arxiv.org/abs/1808.00680} {arXiv:1808.00680} \BibitemShut
  {NoStop}%
\bibitem [{\citenamefont {Wietek}\ \emph {et~al.}(2019)\citenamefont {Wietek},
  \citenamefont {Corboz}, \citenamefont {Wessel}, \citenamefont {Normand},
  \citenamefont {Mila},\ and\ \citenamefont {Honecker}}]{Wietek2019}%
  \BibitemOpen
  \bibfield  {author} {\bibinfo {author} {\bibfnamefont {A.}~\bibnamefont
  {Wietek}}, \bibinfo {author} {\bibfnamefont {P.}~\bibnamefont {Corboz}},
  \bibinfo {author} {\bibfnamefont {S.}~\bibnamefont {Wessel}}, \bibinfo
  {author} {\bibfnamefont {B.}~\bibnamefont {Normand}}, \bibinfo {author}
  {\bibfnamefont {F.}~\bibnamefont {Mila}}, \ and\ \bibinfo {author}
  {\bibfnamefont {A.}~\bibnamefont {Honecker}},\ }\href {\doibase
  10.1103/physrevresearch.1.033038} {\bibfield  {journal} {\bibinfo  {journal}
  {Physical Review Research}\ }\textbf {\bibinfo {volume} {1}},\ \bibinfo
  {pages} {033038} (\bibinfo {year} {2019})},\ \Eprint
  {http://arxiv.org/abs/1907.00008} {arXiv:1907.00008} \BibitemShut {NoStop}%
\bibitem [{\citenamefont {Kshetrimayum}\ \emph {et~al.}(2019)\citenamefont
  {Kshetrimayum}, \citenamefont {Rizzi}, \citenamefont {Eisert},\ and\
  \citenamefont {Or{\'{u}}s}}]{Kshetrimayum2019}%
  \BibitemOpen
  \bibfield  {author} {\bibinfo {author} {\bibfnamefont {A.}~\bibnamefont
  {Kshetrimayum}}, \bibinfo {author} {\bibfnamefont {M.}~\bibnamefont {Rizzi}},
  \bibinfo {author} {\bibfnamefont {J.}~\bibnamefont {Eisert}}, \ and\ \bibinfo
  {author} {\bibfnamefont {R.}~\bibnamefont {Or{\'{u}}s}},\ }\href {\doibase
  10.1103/PhysRevLett.122.070502} {\bibfield  {journal} {\bibinfo  {journal}
  {Physical Review Letters}\ }\textbf {\bibinfo {volume} {122}} (\bibinfo
  {year} {2019}),\ 10.1103/PhysRevLett.122.070502},\ \Eprint
  {http://arxiv.org/abs/1809.08258} {arXiv:1809.08258} \BibitemShut {NoStop}%
\bibitem [{\citenamefont {Qu}\ \emph {et~al.}(2019)\citenamefont {Qu},
  \citenamefont {Li},\ and\ \citenamefont {Xiang}}]{Qu2019}%
  \BibitemOpen
  \bibfield  {author} {\bibinfo {author} {\bibfnamefont {D.~W.}\ \bibnamefont
  {Qu}}, \bibinfo {author} {\bibfnamefont {W.}~\bibnamefont {Li}}, \ and\
  \bibinfo {author} {\bibfnamefont {T.}~\bibnamefont {Xiang}},\ }\href
  {\doibase 10.1103/PhysRevB.100.125121} {\bibfield  {journal} {\bibinfo
  {journal} {Physical Review B}\ }\textbf {\bibinfo {volume} {100}},\ \bibinfo
  {pages} {125121} (\bibinfo {year} {2019})},\ \Eprint
  {http://arxiv.org/abs/1905.12478} {arXiv:1905.12478} \BibitemShut {NoStop}%
\bibitem [{\citenamefont {Czarnik}\ \emph {et~al.}(2012)\citenamefont
  {Czarnik}, \citenamefont {Cincio},\ and\ \citenamefont
  {Dziarmaga}}]{Czarnik2012}%
  \BibitemOpen
  \bibfield  {author} {\bibinfo {author} {\bibfnamefont {P.}~\bibnamefont
  {Czarnik}}, \bibinfo {author} {\bibfnamefont {L.}~\bibnamefont {Cincio}}, \
  and\ \bibinfo {author} {\bibfnamefont {J.}~\bibnamefont {Dziarmaga}},\ }\href
  {\doibase 10.1103/PhysRevB.86.245101} {\bibfield  {journal} {\bibinfo
  {journal} {Physical Review B - Condensed Matter and Materials Physics}\
  }\textbf {\bibinfo {volume} {86}},\ \bibinfo {pages} {245101} (\bibinfo
  {year} {2012})}\BibitemShut {NoStop}%
\bibitem [{\citenamefont {Czarnik}\ and\ \citenamefont
  {Dziarmaga}(2015{\natexlab{a}})}]{Czarnik2015a}%
  \BibitemOpen
  \bibfield  {author} {\bibinfo {author} {\bibfnamefont {P.}~\bibnamefont
  {Czarnik}}\ and\ \bibinfo {author} {\bibfnamefont {J.}~\bibnamefont
  {Dziarmaga}},\ }\href {\doibase 10.1103/PhysRevB.92.035152} {\bibfield
  {journal} {\bibinfo  {journal} {Physical Review B - Condensed Matter and
  Materials Physics}\ }\textbf {\bibinfo {volume} {92}},\ \bibinfo {pages}
  {035152} (\bibinfo {year} {2015}{\natexlab{a}})}\BibitemShut {NoStop}%
\bibitem [{\citenamefont {Czarnik}\ and\ \citenamefont
  {Dziarmaga}(2015{\natexlab{b}})}]{Czarnik2015}%
  \BibitemOpen
  \bibfield  {author} {\bibinfo {author} {\bibfnamefont {P.}~\bibnamefont
  {Czarnik}}\ and\ \bibinfo {author} {\bibfnamefont {J.}~\bibnamefont
  {Dziarmaga}},\ }\href {\doibase 10.1103/PhysRevB.92.035120} {\bibfield
  {journal} {\bibinfo  {journal} {Physical Review B - Condensed Matter and
  Materials Physics}\ }\textbf {\bibinfo {volume} {92}},\ \bibinfo {pages}
  {035120} (\bibinfo {year} {2015}{\natexlab{b}})},\ \Eprint
  {http://arxiv.org/abs/1411.6778} {arXiv:1411.6778} \BibitemShut {NoStop}%
\bibitem [{\citenamefont {Czarnik}\ \emph
  {et~al.}(2019{\natexlab{b}})\citenamefont {Czarnik}, \citenamefont
  {Dziarmaga},\ and\ \citenamefont {Corboz}}]{Czarnik2019a}%
  \BibitemOpen
  \bibfield  {author} {\bibinfo {author} {\bibfnamefont {P.}~\bibnamefont
  {Czarnik}}, \bibinfo {author} {\bibfnamefont {J.}~\bibnamefont {Dziarmaga}},
  \ and\ \bibinfo {author} {\bibfnamefont {P.}~\bibnamefont {Corboz}},\ }\href
  {\doibase 10.1103/PhysRevB.99.035115} {\bibfield  {journal} {\bibinfo
  {journal} {Physical Review B}\ }\textbf {\bibinfo {volume} {99}},\ \bibinfo
  {pages} {035115} (\bibinfo {year} {2019}{\natexlab{b}})}\BibitemShut
  {NoStop}%
\bibitem [{\citenamefont {Kshetrimayum}\ \emph {et~al.}(2017)\citenamefont
  {Kshetrimayum}, \citenamefont {Weimer},\ and\ \citenamefont
  {Or{\'{u}}s}}]{Kshetrimayum2017}%
  \BibitemOpen
  \bibfield  {author} {\bibinfo {author} {\bibfnamefont {A.}~\bibnamefont
  {Kshetrimayum}}, \bibinfo {author} {\bibfnamefont {H.}~\bibnamefont
  {Weimer}}, \ and\ \bibinfo {author} {\bibfnamefont {R.}~\bibnamefont
  {Or{\'{u}}s}},\ }\href {\doibase 10.1038/s41467-017-01511-6} {\bibfield
  {journal} {\bibinfo  {journal} {Nature Communications}\ }\textbf {\bibinfo
  {volume} {8}},\ \bibinfo {pages} {1} (\bibinfo {year} {2017})},\ \Eprint
  {http://arxiv.org/abs/1612.00656} {arXiv:1612.00656} \BibitemShut {NoStop}%
\bibitem [{\citenamefont {Verstraete}\ \emph {et~al.}(2004)\citenamefont
  {Verstraete}, \citenamefont {Garc{\'{i}}a-Ripoll},\ and\ \citenamefont
  {Cirac}}]{Verstraete2004a}%
  \BibitemOpen
  \bibfield  {author} {\bibinfo {author} {\bibfnamefont {F.}~\bibnamefont
  {Verstraete}}, \bibinfo {author} {\bibfnamefont {J.~J.}\ \bibnamefont
  {Garc{\'{i}}a-Ripoll}}, \ and\ \bibinfo {author} {\bibfnamefont {J.~I.}\
  \bibnamefont {Cirac}},\ }\href {\doibase 10.1103/PhysRevLett.93.207204}
  {\bibfield  {journal} {\bibinfo  {journal} {Physical Review Letters}\
  }\textbf {\bibinfo {volume} {93}},\ \bibinfo {pages} {207204} (\bibinfo
  {year} {2004})}\BibitemShut {NoStop}%
\bibitem [{\citenamefont {{Jahromi, Saeed S.,
  Or{\'{u}}s}}(2020)}]{Jahromi2020a}%
  \BibitemOpen
  \bibfield  {author} {\bibinfo {author} {\bibfnamefont {R.}~\bibnamefont
  {{Jahromi, Saeed S., Or{\'{u}}s}}},\ }\href {\doibase
  10.1038/s41598-020-75548-x} {\bibfield  {journal} {\bibinfo  {journal}
  {Scientific Reports}\ }\textbf {\bibinfo {volume} {10}},\ \bibinfo {pages}
  {19051} (\bibinfo {year} {2020})},\ \Eprint {http://arxiv.org/abs/2005.00314}
  {arXiv:2005.00314} \BibitemShut {NoStop}%
\bibitem [{\citenamefont {Altland}\ and\ \citenamefont
  {Zirnbauer}(1997)}]{Altland1997}%
  \BibitemOpen
  \bibfield  {author} {\bibinfo {author} {\bibfnamefont {A.}~\bibnamefont
  {Altland}}\ and\ \bibinfo {author} {\bibfnamefont {M.~R.}\ \bibnamefont
  {Zirnbauer}},\ }\href {\doibase 10.1103/PhysRevB.55.1142} {\bibfield
  {journal} {\bibinfo  {journal} {Physical Review B - Condensed Matter and
  Materials Physics}\ }\textbf {\bibinfo {volume} {55}},\ \bibinfo {pages}
  {1142} (\bibinfo {year} {1997})}\BibitemShut {NoStop}%
\bibitem [{\citenamefont {Or{\'{u}}s}(2014{\natexlab{a}})}]{Orus2014}%
  \BibitemOpen
  \bibfield  {author} {\bibinfo {author} {\bibfnamefont {R.}~\bibnamefont
  {Or{\'{u}}s}},\ }\href {\doibase 10.1016/j.aop.2014.06.013} {\bibfield
  {journal} {\bibinfo  {journal} {Annals of Physics}\ }\textbf {\bibinfo
  {volume} {349}},\ \bibinfo {pages} {117} (\bibinfo {year}
  {2014}{\natexlab{a}})},\ \Eprint {http://arxiv.org/abs/1306.2164}
  {arXiv:1306.2164} \BibitemShut {NoStop}%
\bibitem [{\citenamefont {Or{\'{u}}s}(2014{\natexlab{b}})}]{Orus2014a}%
  \BibitemOpen
  \bibfield  {author} {\bibinfo {author} {\bibfnamefont {R.}~\bibnamefont
  {Or{\'{u}}s}},\ }\href {\doibase 10.1140/epjb/e2014-50502-9} {\bibfield
  {journal} {\bibinfo  {journal} {European Physical Journal B}\ }\textbf
  {\bibinfo {volume} {87}},\ \bibinfo {pages} {280} (\bibinfo {year}
  {2014}{\natexlab{b}})},\ \Eprint {http://arxiv.org/abs/1407.6552}
  {arXiv:1407.6552} \BibitemShut {NoStop}%
\bibitem [{\citenamefont {Or{\'{u}}s}(2019)}]{Orus2019}%
  \BibitemOpen
  \bibfield  {author} {\bibinfo {author} {\bibfnamefont {R.}~\bibnamefont
  {Or{\'{u}}s}},\ }\href {\doibase 10.1038/s42254-019-0086-7} {\bibfield
  {journal} {\bibinfo  {journal} {Nature Reviews Physics}\ }\textbf {\bibinfo
  {volume} {1}},\ \bibinfo {pages} {538} (\bibinfo {year} {2019})}\BibitemShut
  {NoStop}%
\bibitem [{\citenamefont {Ran}\ \emph {et~al.}(2017)\citenamefont {Ran},
  \citenamefont {Tirrito}, \citenamefont {Peng}, \citenamefont {Chen},
  \citenamefont {Su},\ and\ \citenamefont {Lewenstein}}]{Ran2017}%
  \BibitemOpen
  \bibfield  {author} {\bibinfo {author} {\bibfnamefont {S.-J.}\ \bibnamefont
  {Ran}}, \bibinfo {author} {\bibfnamefont {E.}~\bibnamefont {Tirrito}},
  \bibinfo {author} {\bibfnamefont {C.}~\bibnamefont {Peng}}, \bibinfo {author}
  {\bibfnamefont {X.}~\bibnamefont {Chen}}, \bibinfo {author} {\bibfnamefont
  {G.}~\bibnamefont {Su}}, \ and\ \bibinfo {author} {\bibfnamefont
  {M.}~\bibnamefont {Lewenstein}},\ }\href {\doibase
  10.1016/j.hrmr.2011.11.009} {\  (\bibinfo {year} {2017}),\
  10.1016/j.hrmr.2011.11.009},\ \Eprint {http://arxiv.org/abs/1708.09213}
  {arXiv:1708.09213} \BibitemShut {NoStop}%
\bibitem [{\citenamefont {Biamonte}\ and\ \citenamefont
  {Bergholm}(2017)}]{Biamonte2017}%
  \BibitemOpen
  \bibfield  {author} {\bibinfo {author} {\bibfnamefont {J.}~\bibnamefont
  {Biamonte}}\ and\ \bibinfo {author} {\bibfnamefont {V.}~\bibnamefont
  {Bergholm}},\ }\href {http://arxiv.org/abs/1708.00006} {\  (\bibinfo {year}
  {2017})},\ \Eprint {http://arxiv.org/abs/1708.00006} {arXiv:1708.00006}
  \BibitemShut {NoStop}%
\bibitem [{\citenamefont {Verstraete}\ \emph {et~al.}(2008)\citenamefont
  {Verstraete}, \citenamefont {Murg},\ and\ \citenamefont
  {Cirac}}]{Verstraete2008}%
  \BibitemOpen
  \bibfield  {author} {\bibinfo {author} {\bibfnamefont {F.}~\bibnamefont
  {Verstraete}}, \bibinfo {author} {\bibfnamefont {V.}~\bibnamefont {Murg}}, \
  and\ \bibinfo {author} {\bibfnamefont {J.~I.}\ \bibnamefont {Cirac}},\ }\href
  {\doibase 10.1080/14789940801912366} {\bibfield  {journal} {\bibinfo
  {journal} {Advances in Physics}\ }\textbf {\bibinfo {volume} {57}},\ \bibinfo
  {pages} {143} (\bibinfo {year} {2008})},\ \Eprint
  {http://arxiv.org/abs/0907.2796} {arXiv:0907.2796} \BibitemShut {NoStop}%
\bibitem [{\citenamefont {White}(1993)}]{White1993}%
  \BibitemOpen
  \bibfield  {author} {\bibinfo {author} {\bibfnamefont {S.~R.}\ \bibnamefont
  {White}},\ }\href {\doibase 10.1103/PhysRevB.48.10345} {\bibfield  {journal}
  {\bibinfo  {journal} {Physical Review B}\ }\textbf {\bibinfo {volume} {48}},\
  \bibinfo {pages} {10345} (\bibinfo {year} {1993})}\BibitemShut {NoStop}%
\bibitem [{\citenamefont {White}\ and\ \citenamefont
  {Feiguin}(2004)}]{White2004}%
  \BibitemOpen
  \bibfield  {author} {\bibinfo {author} {\bibfnamefont {S.~R.}\ \bibnamefont
  {White}}\ and\ \bibinfo {author} {\bibfnamefont {A.~E.}\ \bibnamefont
  {Feiguin}},\ }\href {\doibase 10.1103/PhysRevLett.93.076401} {\bibfield
  {journal} {\bibinfo  {journal} {Physical Review Letters}\ }\textbf {\bibinfo
  {volume} {93}},\ \bibinfo {pages} {076401} (\bibinfo {year} {2004})},\
  \Eprint {http://arxiv.org/abs/0403310} {arXiv:0403310 [cond-mat]}
  \BibitemShut {NoStop}%
\bibitem [{\citenamefont {Verstraete}\ \emph {et~al.}(2006)\citenamefont
  {Verstraete}, \citenamefont {Wolf}, \citenamefont {Perez-Garcia},\ and\
  \citenamefont {Cirac}}]{Verstraete2006}%
  \BibitemOpen
  \bibfield  {author} {\bibinfo {author} {\bibfnamefont {F.}~\bibnamefont
  {Verstraete}}, \bibinfo {author} {\bibfnamefont {M.~M.}\ \bibnamefont
  {Wolf}}, \bibinfo {author} {\bibfnamefont {D.}~\bibnamefont {Perez-Garcia}},
  \ and\ \bibinfo {author} {\bibfnamefont {J.~I.}\ \bibnamefont {Cirac}},\
  }\href {\doibase 10.1103/PhysRevLett.96.220601} {\bibfield  {journal}
  {\bibinfo  {journal} {Physical Review Letters}\ }\textbf {\bibinfo {volume}
  {96}},\ \bibinfo {pages} {220601} (\bibinfo {year} {2006})},\ \Eprint
  {http://arxiv.org/abs/0601075} {arXiv:0601075 [quant-ph]} \BibitemShut
  {NoStop}%
\bibitem [{\citenamefont {Corboz}(2016)}]{Corboz2016}%
  \BibitemOpen
  \bibfield  {author} {\bibinfo {author} {\bibfnamefont {P.}~\bibnamefont
  {Corboz}},\ }\href {\doibase 10.1103/PhysRevB.94.035133} {\bibfield
  {journal} {\bibinfo  {journal} {Physical Review B}\ }\textbf {\bibinfo
  {volume} {94}},\ \bibinfo {pages} {035133} (\bibinfo {year} {2016})},\
  \Eprint {http://arxiv.org/abs/1605.03006} {arXiv:1605.03006} \BibitemShut
  {NoStop}%
\bibitem [{\citenamefont {Or{\'{u}}s}\ and\ \citenamefont
  {Vidal}(2009)}]{Orus2009}%
  \BibitemOpen
  \bibfield  {author} {\bibinfo {author} {\bibfnamefont {R.}~\bibnamefont
  {Or{\'{u}}s}}\ and\ \bibinfo {author} {\bibfnamefont {G.}~\bibnamefont
  {Vidal}},\ }\href {\doibase 10.1103/PhysRevB.80.094403} {\bibfield  {journal}
  {\bibinfo  {journal} {Physical Review B - Condensed Matter and Materials
  Physics}\ }\textbf {\bibinfo {volume} {80}},\ \bibinfo {pages} {094403}
  (\bibinfo {year} {2009})},\ \Eprint {http://arxiv.org/abs/0905.3225}
  {arXiv:0905.3225} \BibitemShut {NoStop}%
\bibitem [{\citenamefont {Phien}\ \emph {et~al.}(2015)\citenamefont {Phien},
  \citenamefont {Bengua}, \citenamefont {Tuan}, \citenamefont {Corboz},\ and\
  \citenamefont {Or{\'{u}}s}}]{Phien2015}%
  \BibitemOpen
  \bibfield  {author} {\bibinfo {author} {\bibfnamefont {H.~N.}\ \bibnamefont
  {Phien}}, \bibinfo {author} {\bibfnamefont {J.~A.}\ \bibnamefont {Bengua}},
  \bibinfo {author} {\bibfnamefont {H.~D.}\ \bibnamefont {Tuan}}, \bibinfo
  {author} {\bibfnamefont {P.}~\bibnamefont {Corboz}}, \ and\ \bibinfo {author}
  {\bibfnamefont {R.}~\bibnamefont {Or{\'{u}}s}},\ }\href {\doibase
  10.1103/PhysRevB.92.035142} {\bibfield  {journal} {\bibinfo  {journal}
  {Physical Review B - Condensed Matter and Materials Physics}\ }\textbf
  {\bibinfo {volume} {92}},\ \bibinfo {pages} {035142} (\bibinfo {year}
  {2015})},\ \Eprint {http://arxiv.org/abs/1503.05345} {arXiv:1503.05345}
  \BibitemShut {NoStop}%
\bibitem [{\citenamefont {Jordan}\ \emph {et~al.}(2008)\citenamefont {Jordan},
  \citenamefont {Or{\'{u}}s}, \citenamefont {Vidal}, \citenamefont
  {Verstraete},\ and\ \citenamefont {Cirac}}]{Vidal2007}%
  \BibitemOpen
  \bibfield  {author} {\bibinfo {author} {\bibfnamefont {J.}~\bibnamefont
  {Jordan}}, \bibinfo {author} {\bibfnamefont {R.}~\bibnamefont {Or{\'{u}}s}},
  \bibinfo {author} {\bibfnamefont {G.}~\bibnamefont {Vidal}}, \bibinfo
  {author} {\bibfnamefont {F.}~\bibnamefont {Verstraete}}, \ and\ \bibinfo
  {author} {\bibfnamefont {J.~I.}\ \bibnamefont {Cirac}},\ }\href {\doibase
  10.1103/PhysRevLett.101.250602} {\bibfield  {journal} {\bibinfo  {journal}
  {Physical Review Letters}\ }\textbf {\bibinfo {volume} {101}},\ \bibinfo
  {pages} {250602} (\bibinfo {year} {2008})},\ \Eprint
  {http://arxiv.org/abs/0703788} {arXiv:0703788 [cond-mat]} \BibitemShut
  {NoStop}%
\bibitem [{\citenamefont {Levin}\ and\ \citenamefont {Nave}(2007)}]{Levin2007}%
  \BibitemOpen
  \bibfield  {author} {\bibinfo {author} {\bibfnamefont {M.}~\bibnamefont
  {Levin}}\ and\ \bibinfo {author} {\bibfnamefont {C.~P.}\ \bibnamefont
  {Nave}},\ }\href {\doibase 10.1103/PhysRevLett.99.120601} {\bibfield
  {journal} {\bibinfo  {journal} {Physical Review Letters}\ }\textbf {\bibinfo
  {volume} {99}},\ \bibinfo {pages} {120601} (\bibinfo {year} {2007})},\
  \Eprint {http://arxiv.org/abs/0611687} {arXiv:0611687 [cond-mat]}
  \BibitemShut {NoStop}%
\bibitem [{\citenamefont {Baskaran}\ \emph {et~al.}(2007)\citenamefont
  {Baskaran}, \citenamefont {Mandal},\ and\ \citenamefont
  {Shankar}}]{Baskaran2007}%
  \BibitemOpen
  \bibfield  {author} {\bibinfo {author} {\bibfnamefont {G.}~\bibnamefont
  {Baskaran}}, \bibinfo {author} {\bibfnamefont {S.}~\bibnamefont {Mandal}}, \
  and\ \bibinfo {author} {\bibfnamefont {R.}~\bibnamefont {Shankar}},\ }\href
  {\doibase 10.1103/PhysRevLett.98.247201} {\bibfield  {journal} {\bibinfo
  {journal} {Physical Review Letters}\ }\textbf {\bibinfo {volume} {98}},\
  \bibinfo {pages} {247201} (\bibinfo {year} {2007})},\ \Eprint
  {http://arxiv.org/abs/0611547} {arXiv:0611547 [cond-mat]} \BibitemShut
  {NoStop}%
\bibitem [{\citenamefont {Luttinger}(1964)}]{Luttinger1964}%
  \BibitemOpen
  \bibfield  {author} {\bibinfo {author} {\bibfnamefont {J.~M.}\ \bibnamefont
  {Luttinger}},\ }\href {\doibase 10.1103/PhysRev.135.A1505} {\bibfield
  {journal} {\bibinfo  {journal} {Physical Review}\ }\textbf {\bibinfo {volume}
  {135}},\ \bibinfo {pages} {A1505} (\bibinfo {year} {1964})}\BibitemShut
  {NoStop}%
\bibitem [{\citenamefont {Qin}\ \emph {et~al.}(2011)\citenamefont {Qin},
  \citenamefont {Niu},\ and\ \citenamefont {Shi}}]{Qin2011}%
  \BibitemOpen
  \bibfield  {author} {\bibinfo {author} {\bibfnamefont {T.}~\bibnamefont
  {Qin}}, \bibinfo {author} {\bibfnamefont {Q.}~\bibnamefont {Niu}}, \ and\
  \bibinfo {author} {\bibfnamefont {J.}~\bibnamefont {Shi}},\ }\href {\doibase
  10.1103/PhysRevLett.107.236601} {\bibfield  {journal} {\bibinfo  {journal}
  {Physical Review Letters}\ }\textbf {\bibinfo {volume} {107}},\ \bibinfo
  {pages} {236601} (\bibinfo {year} {2011})}\BibitemShut {NoStop}%
\bibitem [{\citenamefont {Matsumoto}\ and\ \citenamefont
  {Murakami}(2011)}]{Matsumoto2011}%
  \BibitemOpen
  \bibfield  {author} {\bibinfo {author} {\bibfnamefont {R.}~\bibnamefont
  {Matsumoto}}\ and\ \bibinfo {author} {\bibfnamefont {S.}~\bibnamefont
  {Murakami}},\ }\href {\doibase 10.1103/PhysRevB.84.184406} {\bibfield
  {journal} {\bibinfo  {journal} {Physical Review B - Condensed Matter and
  Materials Physics}\ }\textbf {\bibinfo {volume} {84}},\ \bibinfo {pages}
  {184406} (\bibinfo {year} {2011})},\ \Eprint {http://arxiv.org/abs/1106.1987}
  {arXiv:1106.1987} \BibitemShut {NoStop}%
\bibitem [{\citenamefont {Go}\ \emph {et~al.}(2019)\citenamefont {Go},
  \citenamefont {Jung},\ and\ \citenamefont {Moon}}]{Go2019}%
  \BibitemOpen
  \bibfield  {author} {\bibinfo {author} {\bibfnamefont {A.}~\bibnamefont
  {Go}}, \bibinfo {author} {\bibfnamefont {J.}~\bibnamefont {Jung}}, \ and\
  \bibinfo {author} {\bibfnamefont {E.~G.}\ \bibnamefont {Moon}},\ }\href
  {\doibase 10.1103/PhysRevLett.122.147203} {\bibfield  {journal} {\bibinfo
  {journal} {Physical Review Letters}\ }\textbf {\bibinfo {volume} {122}},\
  \bibinfo {pages} {147203} (\bibinfo {year} {2019})},\ \Eprint
  {http://arxiv.org/abs/1808.09457} {arXiv:1808.09457} \BibitemShut {NoStop}%
\bibitem [{\citenamefont {Meng}\ and\ \citenamefont
  {Balents}(2012)}]{Meng2012}%
  \BibitemOpen
  \bibfield  {author} {\bibinfo {author} {\bibfnamefont {T.}~\bibnamefont
  {Meng}}\ and\ \bibinfo {author} {\bibfnamefont {L.}~\bibnamefont {Balents}},\
  }\href {\doibase 10.1103/PhysRevB.86.054504} {\bibfield  {journal} {\bibinfo
  {journal} {Physical Review B - Condensed Matter and Materials Physics}\
  }\textbf {\bibinfo {volume} {86}},\ \bibinfo {pages} {054504} (\bibinfo
  {year} {2012})},\ \Eprint {http://arxiv.org/abs/1205.5202} {arXiv:1205.5202}
  \BibitemShut {NoStop}%
\bibitem [{\citenamefont {Yoshioka}\ \emph {et~al.}(2018)\citenamefont
  {Yoshioka}, \citenamefont {Imai},\ and\ \citenamefont
  {Sigrist}}]{Yoshioka2018}%
  \BibitemOpen
  \bibfield  {author} {\bibinfo {author} {\bibfnamefont {N.}~\bibnamefont
  {Yoshioka}}, \bibinfo {author} {\bibfnamefont {Y.}~\bibnamefont {Imai}}, \
  and\ \bibinfo {author} {\bibfnamefont {M.}~\bibnamefont {Sigrist}},\ }\href
  {\doibase 10.7566/JPSJ.87.124602} {\bibfield  {journal} {\bibinfo  {journal}
  {Journal of the Physical Society of Japan}\ }\textbf {\bibinfo {volume} {87}}
  (\bibinfo {year} {2018}),\ 10.7566/JPSJ.87.124602},\ \Eprint
  {http://arxiv.org/abs/1804.03843} {arXiv:1804.03843} \BibitemShut {NoStop}%
\bibitem [{\citenamefont {Gao}\ \emph {et~al.}(2019)\citenamefont {Gao},
  \citenamefont {Hickey}, \citenamefont {Xiang}, \citenamefont {Trebst},\ and\
  \citenamefont {Chen}}]{Gao2019}%
  \BibitemOpen
  \bibfield  {author} {\bibinfo {author} {\bibfnamefont {Y.~H.}\ \bibnamefont
  {Gao}}, \bibinfo {author} {\bibfnamefont {C.}~\bibnamefont {Hickey}},
  \bibinfo {author} {\bibfnamefont {T.}~\bibnamefont {Xiang}}, \bibinfo
  {author} {\bibfnamefont {S.}~\bibnamefont {Trebst}}, \ and\ \bibinfo {author}
  {\bibfnamefont {G.}~\bibnamefont {Chen}},\ }\href {\doibase
  10.1103/physrevresearch.1.013014} {\enquote {\bibinfo {title} {{Thermal Hall
  signatures of non-Kitaev spin liquids in honeycomb Kitaev materials}},}\ }
  (\bibinfo {year} {2019}),\ \Eprint {http://arxiv.org/abs/1905.11321}
  {arXiv:1905.11321} \BibitemShut {NoStop}%
\bibitem [{\citenamefont {Kimchi}\ \emph {et~al.}(2014)\citenamefont {Kimchi},
  \citenamefont {Analytis},\ and\ \citenamefont {Vishwanath}}]{Kimchi2014}%
  \BibitemOpen
  \bibfield  {author} {\bibinfo {author} {\bibfnamefont {I.}~\bibnamefont
  {Kimchi}}, \bibinfo {author} {\bibfnamefont {J.~G.}\ \bibnamefont
  {Analytis}}, \ and\ \bibinfo {author} {\bibfnamefont {A.}~\bibnamefont
  {Vishwanath}},\ }\href {\doibase 10.1103/PhysRevB.90.205126} {\bibfield
  {journal} {\bibinfo  {journal} {Physical Review B - Condensed Matter and
  Materials Physics}\ }\textbf {\bibinfo {volume} {90}},\ \bibinfo {pages}
  {205126} (\bibinfo {year} {2014})},\ \Eprint {http://arxiv.org/abs/1309.1171}
  {arXiv:1309.1171} \BibitemShut {NoStop}%
\bibitem [{\citenamefont {Bruognolo}\ \emph {et~al.}(2017)\citenamefont
  {Bruognolo}, \citenamefont {Zhu}, \citenamefont {White},\ and\ \citenamefont
  {Stoudenmire}}]{Bruognolo2017}%
  \BibitemOpen
  \bibfield  {author} {\bibinfo {author} {\bibfnamefont {B.}~\bibnamefont
  {Bruognolo}}, \bibinfo {author} {\bibfnamefont {Z.}~\bibnamefont {Zhu}},
  \bibinfo {author} {\bibfnamefont {S.~R.}\ \bibnamefont {White}}, \ and\
  \bibinfo {author} {\bibfnamefont {E.~M.}\ \bibnamefont {Stoudenmire}},\
  }\href {http://arxiv.org/abs/1705.05578} {\  (\bibinfo {year} {2017})},\
  \Eprint {http://arxiv.org/abs/1705.05578} {arXiv:1705.05578} \BibitemShut
  {NoStop}%
\bibitem [{\citenamefont {Chen}\ and\ \citenamefont
  {Stoudenmire}(2020)}]{Chen2020}%
  \BibitemOpen
  \bibfield  {author} {\bibinfo {author} {\bibfnamefont {J.}~\bibnamefont
  {Chen}}\ and\ \bibinfo {author} {\bibfnamefont {E.~M.}\ \bibnamefont
  {Stoudenmire}},\ }\href {\doibase 10.1103/PhysRevB.101.195119} {\bibfield
  {journal} {\bibinfo  {journal} {Physical Review B}\ }\textbf {\bibinfo
  {volume} {101}},\ \bibinfo {pages} {195119} (\bibinfo {year} {2020})},\
  \Eprint {http://arxiv.org/abs/1910.09142} {arXiv:1910.09142} \BibitemShut
  {NoStop}%
\end{thebibliography}%
\end{document}